\documentclass[aps,prl,twocolumn,showpacs,superscriptaddress,footinbib]{revtex4}  % for review and submission
\usepackage{graphicx}  % needed for figures
\usepackage{dcolumn}   % needed for some tables
\usepackage{bm}        % for math
\usepackage{braket}
\usepackage{exscale}                  % correct scaling of math-symbols
\usepackage[intlimits]{amsmath}       % improved mathematical typesetting
\usepackage{amsfonts}
\usepackage{amssymb,amscd}
\usepackage{color}
\usepackage{hyperref}
\usepackage{subfigure}
\usepackage[normalem]{ulem}
\usepackage{slashed}
\usepackage{leftidx}

\graphicspath{{Plots/}}

\definecolor{mygreen}{rgb}{0,0.5,0}
\definecolor{myblue}{rgb}{0,0,0.75}
\definecolor{mymagenta}{cmyk}{0,1,0,0.12}
\definecolor{mygray}{rgb}{0.5,0.5,0.5}

\newcommand{\Fig}[1]{Fig.~\ref{#1}}

\usepackage{umoline}

\begin{document}

\title{Non-cancellation of the parity anomaly in the strong-field regime of QED$_{2+1}$}

\author{R.~Ott}
\email[]{ott@thphys.uni-heidelberg.de}
\affiliation{Heidelberg University, Institut f\"{u}r Theoretische Physik, Philosophenweg 16, 69120 Heidelberg, Germany}

\author{T.~V.~Zache}
\affiliation{Heidelberg University, Institut f\"{u}r Theoretische Physik, Philosophenweg 16, 69120 Heidelberg, Germany}

\author{N.~Mueller}
\affiliation{Physics Department, Brookhaven National Laboratory, Building 510A, Upton, New York 11973, USA}

\author{J.~Berges}
\affiliation{Heidelberg University, Institut f\"{u}r Theoretische Physik, Philosophenweg 16, 69120 Heidelberg, Germany}

\begin{abstract}
Quantum fluctuations lead to an anomalous violation of parity symmetry in quantum electrodynamics for an even number of spatial dimensions. While the leading parity-odd electric current vanishes in vacuum, we uncover a non-cancellation of the anomaly for strong electric fields with distinct macroscopic signatures. We perform real-time lattice simulations with fully dynamical gauge fields and Wilson fermions in $2+1$ space-time dimensions. In the static field limit, relevant at early times, we solve the problem analytically. Our results point out the fundamental role of quantum anomalies for strong-field phenomena, relevant for a wide range of condensed matter and high-energy applications, but also for the next generation of gauge theory quantum simulators.
\end{abstract}
\maketitle

{\it Introduction.}
Quantum electrodynamics (QED) is well understood in vacuum and for weak electromagnetic fields. Much less is known in the strong-field regime relevant for a wide range of applications in condensed matter physics~\cite{Q.Lietal., J.Xiongetal.,Hosur:2013kxa,Gorbar:2017lnp}, relativistic nuclear collision experiments~\cite{Kharzeev:2007jp,Fukushima:2008xe,Kharzeev:2015znc,Skokov:2016yrj}, precision spectroscopy of highly charged ions~\cite{BLAUM20061,BLAUMNature}, or future laser facilities~\cite{DiPiazza:2011tq}. A typical electric field strength characterizing the strong-field regime is $E_c = m^2/e$, where $m$ denotes the electron mass and $e$ is the gauge coupling~\footnote{We use natural units with the speed of light $c$ and the reduced Planck constant $\hbar$ being unity.}. For fields exceeding $E_c$, the vacuum of QED becomes unstable against Schwinger pair production of electron-positron pairs~\cite{PhysRev.82.664}.  

In recent years, strong-field QED phenomena have also become a focus of research for quantum simulations~\cite{Wiese:2013uua,Zohar:2015hwa,CiracZoller,Hauke_2012,Tagliacozzo:2012df,Tagliacozzo:2014bta,Haegeman:2014maa,Lamm:2019bik}. While the first implementation using trapped ions still concerned the weak-field limit of QED in one spatial dimension~\cite{BlattZoller}, the next generation of experiments aims at setups in 2+1 space-time dimensions \cite{Zohar:2015hwa,PhysRevLett.118.070501,Zohar:2016iic} and with strong fields~\cite{Pichler:2015yqa,Hebenstreit:2013baa,Zache:2018jbt}. Among the most intriguing phenomena that will become accessible is the anomalous violation of parity by quantum fluctuations in QED for an even number of spatial dimensions~\cite{COSTE1989631,PhysRevD.29.2366,Redlich:1983kn,PhysRevLett.51.2077,Karthik:2018tnh,Karthik:2016ppr,Lapa:2019fiv,Kapustin:2014lwa,Deser:1997nv,Nogueira:2005aj,Callan:1984sa,Steinberg:2019uqb}.  However, there is a cancellation of the anomalous electric current with a parity-odd contribution induced by the fermion mass such that the phenomenon is suppressed for weak fields in vacuum~\cite{COSTE1989631,PhysRevD.29.2366}.    

In this work, we establish a non-cancellation of the parity anomaly in the strong-field regime. This is shown to have dramatic consequences for the presence of anomalous currents with distinctive macroscopic signatures. The net parity-odd electric currents change the macroscopic gauge field evolution and lead to a dynamical rotation of the electric field vector. To validate this discovery, we investigate the fundamental processes both analytically, as well as numerically using real-time lattice simulations with second-order improved Wilson fermions~\cite{Mueller:2016ven,Spitz:2018eps}. The lattice simulations employ classical-statistical reweighting techniques for the bosonic gauge fields, which accurately describe the dynamics in the strong-field regime including the mutual back-action of the induced currents and the applied fields~\cite{Aarts:1998td,Berges:2010zv,Saffin:2011kc,Hebenstreit:2013qxa,Kasper:2014uaa,Buividovich:2015jfa,Gelfand:2016prm,Tanji:2016dka,Mueller:2016ven,Mueller:2016aao}. Our analytical results neglect back-action and apply only to not too late times. In their range of validity they are found to describe the simulation data well.    

\begin{figure*}[!tbp]
	\def\svgwidth{510pt}
	\centering
	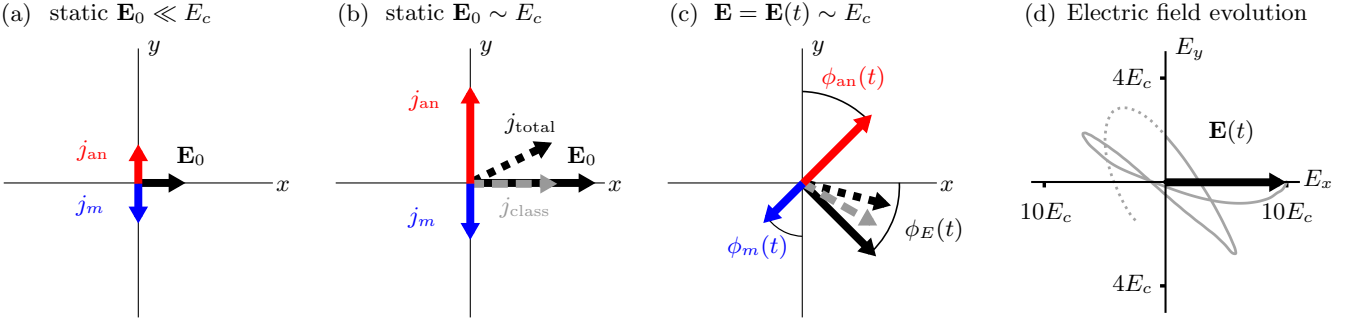
	\caption{{\bf Overview:} Panel \textbf{(a)} illustrates the weak-field limit, where parity-odd currents from the mass term, $j_m$, and the quantum anomaly, $j_{\text{an}}$, cancel. Panel \textbf{(b)} shows the situation for homogeneous fields with strengths beyond the Schwinger limit $E_c$ with current $j_{\text{class}}$ from pair production. In this case the total current is not aligned with the applied field $\mathbf{E}_0$ because of parity non-cancellation. Including the mutual back-action of fermions and gauge fields all components rotate, as pictured in panel \textbf{(c)}. The full evolution of the electric field $\mathbf{E}(t)$ at times $t \le 12/m $ (solid line) and beyond (dotted) is shown in panel \textbf{(d)}.}
	\label{fig:Schema}
\end{figure*}

{\it Parity anomaly cancellation in weak fields.}
In two spatial dimensions, QED breaks space-inversion symmetry (parity) both by a mass term $\sim m$ in the Dirac equation, 
\begin{equation}
\label{eq:Dirac-eq}
\left(i \gamma^{\mu}\partial_{\mu}  -  e \gamma^{\mu} \hat{A}_{\mu}(\mathbf{x},t) -m \right) \hat{\psi}^{\alpha}(\mathbf{x},t) = 0 \; , 
\end{equation}
as well as by a quantum anomaly~\cite{COSTE1989631,PhysRevD.29.2366}. The  index $\alpha = 1,\ldots, N_f$ labels flavors of the fermion field operator $\hat{\psi}^\alpha$, with $N_f=1$ for QED. Here, $\hat{A}_{\mu}$ is the gauge potential, $e$ the coupling, and summation over repeated indices is implied. For the representation of the Dirac matrices $\gamma^{\mu}$ with $\mu = 0,1,2$ we choose the Pauli matrices $(\sigma_{3},i\sigma_{1},-i\sigma_{2})$.
The gauge field operator, in terms of the gauge-invariant field strength tensor $\hat{F}_{\nu\rho}=\partial_\nu \hat{A}_\rho-\partial_\rho \hat{A}_\nu$, evolves as
\begin{equation}
\label{eq:Gauge-field-equation}
\partial_{\mu} \hat{F}^{\mu\nu}(\mathbf{x},t) = - \frac{e}{2} \;\mathrm{tr}\left\{\gamma^{\nu}\left[\hat{\psi}^{\alpha}(\mathbf{x},t),\hat{\bar{\psi}}^{\alpha}(\mathbf{x},t)\right]\right\} \; ,
\end{equation}
where $\hat{\bar{\psi}}^{\alpha} \equiv \hat{\psi}^{\alpha\dagger} \gamma^{0}$. Here $[\cdot,\cdot]$ is the commutator and $\mathrm{tr}\{\cdot\}$ denotes a trace over flavor and spinor components.

Parity violation manifests itself as a non-zero (parity-odd) expectation value of the fermion electric current $j^\mu_{\text{total}} = \mathrm{tr} \{\gamma^\mu\langle  [\hat{\psi}^{\alpha}, \hat{\bar{\psi}}^{\alpha}]\rangle \}/2$. In vacuum, this current can be decomposed into a mass-related part, $j^\mu_{\text{m}}$, and a part originating from the anomaly, $j^\mu_{\text{an}}$. In the presence of a weak, slowly evolving homogeneous background gauge field the leading contribution to $j^\mu_{\text{m}}$ is given by~\cite{COSTE1989631,PhysRevD.29.2366} 
\begin{equation}\label{eq:groundstatemass}
j^\mu_{\text{m}}(t) \stackrel{\text{weak field}}{=} -\frac{m}{|m|}  \frac{e}{8\pi}  \epsilon^{\mu\nu\rho}F_{\nu\rho}(t) 
\end{equation}
with $F_{\nu\rho} = \langle\hat{F}_{\nu\rho}\rangle$, and the anomaly contribution reads
\begin{align}\label{eq:j-odd}
j^\mu_{\text{an}}(t) =  \frac{e}{8\pi}  \epsilon^{\mu\nu\rho}F_{\nu\rho}(t) \, .
\end{align}
Both contributions have equal magnitude and, as indicated in panel (a) of \Fig{fig:Schema}, they cancel each other for $m > 0$, resulting in an overall parity conservation with $j^\mu_{\text{total}} = j^\mu_{\text{m}} + j^\mu_{\text{an}} = 0$ in this case~\footnote{A similar cancellation of anomalous contributions appears for the Adler-Bell-Jackiw anomaly in $1+1$ space-time dimensions for non-zero fermion mass and weak electric field $E \ll E_c$ \cite{Ambjorn:1983hp}.}.

\textit{Anomaly non-cancellation in strong background fields.}
To extend these results to the strong-field regime, we first consider a static homogeneous electric field $\mathbf{E}_0 = \begin{pmatrix}E_x , 0\end{pmatrix}$. This turns out to be analytically tractable also in the presence of field strengths beyond the Schwinger limit. We solve the Hamiltonian operator equation (\ref{eq:Dirac-eq}) for an infinite volume in temporal-axial gauge, with $A_0 = 0$ and $\mathbf{A} = -\mathbf{E}_0 t$, for fermion vacuum initial conditions prepared in the remote past \cite{Hebenstreit:2011pm}. We then construct currents from bilinears of fermionic field operators in a standard mode decomposition. 

In this setup, the total current includes both parity-even and parity-odd contributions, as indicated in \Fig{fig:Schema}, panel (b).
The parity-even current $j_{\mathrm{class}}$, pointing along the $x$-direction, is due to conventional Schwinger pair production for strong fields, ubiquitous in one \cite{Hebenstreit:2013qxa}, two (this work) and three \cite{Kasper:2014uaa,Mueller:2016aao} spatial dimensions.
Conversely, the parity-odd current component along the $y$-direction represents the sum of the contribution from the mass term, $j^{y}_{\text{m}}$, and the anomalous current, $j^{y}_{\text{an}}$.

We find that $j^{y}_{\text{m}}$ receives two separate contributions of distinct physical origin, described by an integral over momenta in $y$-direction:   
\begin{equation}
\label{eq:Erf-two-terms}
	j^{y}_{\text{m}} = - \frac{e E_x m}{4 \pi} \int_{-\infty}^{\infty}\! \frac{\mathrm{d} p_y}{(2\pi)} \left( \frac{1}{p_y^2+m^2} - \frac{e^{-\pi \frac{p_y^2+m^2}{eE_x}}}{p_y^2+m^2} \right)  .
\end{equation}
The first term in the integrand of Eq.~(\ref{eq:Erf-two-terms}) represents the known vacuum contribution~\cite{COSTE1989631,PhysRevD.29.2366}, which reads
\begin{equation}
	-\frac{e E_x m}{4 \pi} \int_{-\infty}^{\infty}\! \frac{\mathrm{d} p_y}{(2\pi)}\,  \frac{1}{p_y^2+m^2}  = - \frac{m}{|m|} \frac{eE_x}{4\pi} \; 
\end{equation}
and agrees with Eq.~(\ref{eq:groundstatemass}).  The second term in Eq.~(\ref{eq:Erf-two-terms}) represents a medium correction, which is seen to be only suppressed below the critical field strength for Schwinger pair production with rate~$\sim \exp(-\pi E_c/E_x)$.  

Taking both vacuum and medium contributions into account, Eq.~(\ref{eq:Erf-two-terms}) reads
\begin{align} \label{eq:j-odd-Erf}
	j^{y}_{\text{m}} = - \frac{m}{|m|} \frac{eE_x}{4\pi}  \; \text{Erf}\bigg(\sqrt{\frac{\pi m^2}{eE_x}}\bigg) \, .
\end{align}
For weak fields Eq.~(\ref{eq:groundstatemass}) is recovered, whereas for fields exceeding $E_c$ the modulus of $j^{y}_{\text{m}}$ can be significantly reduced. In contrast, the anomalous contribution $j^{y}_{\text{an}} =  eE_x/(4\pi)$ is unchanged beyond the Schwinger limit and agrees with Eq.~(\ref{eq:j-odd}), which we verified explicity using a gauge invariant (Pauli-Villars) regularization scheme~\cite{RevModPhys.21.434}. As a consequence, in the strong-field regime $j^{y}_{\text{m}}$ and $j^{y}_{\text{an}}$ no longer cancel. 

\begin{figure}
	\flushleft
	\includegraphics[width=0.473\textwidth]{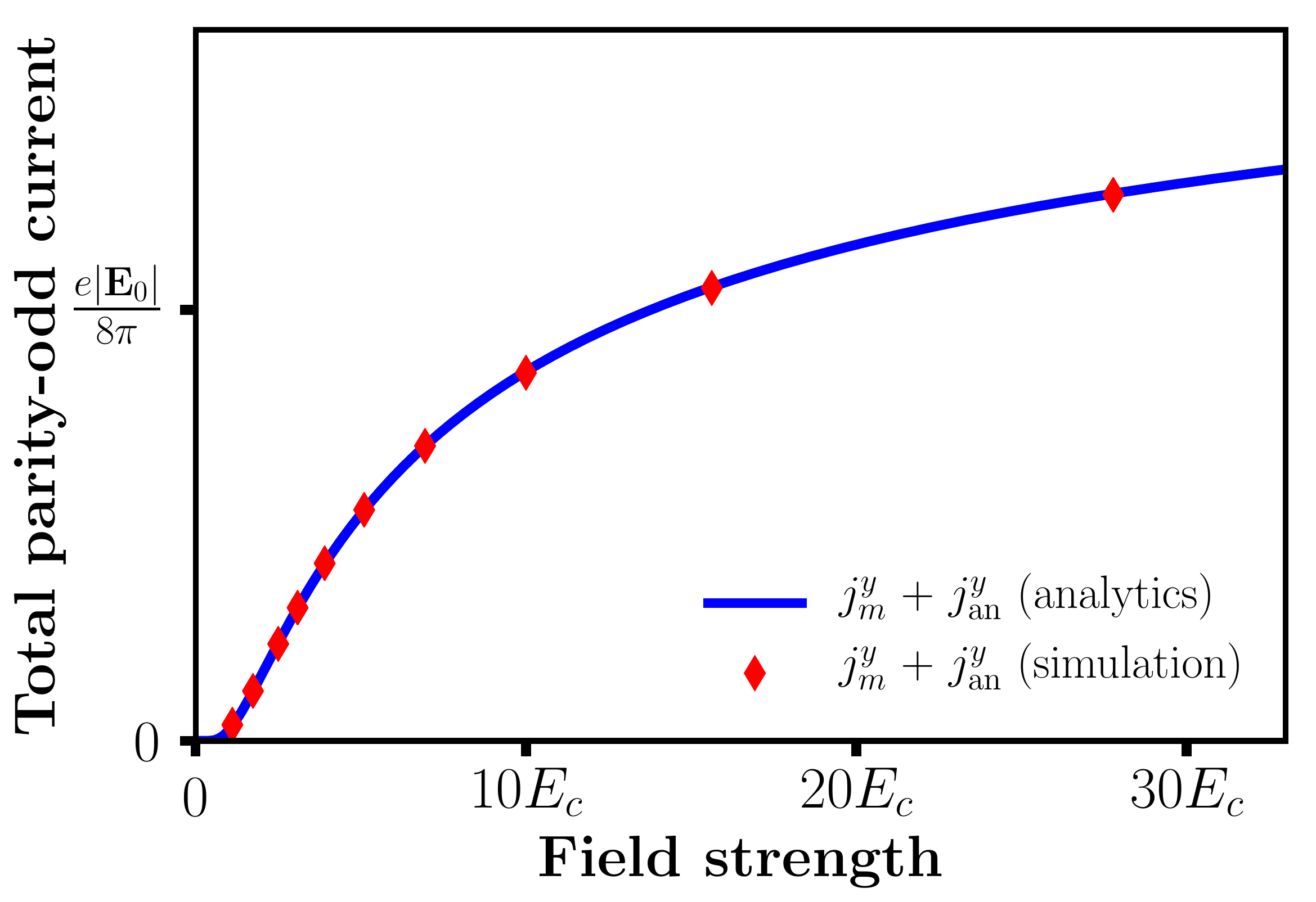}
	\caption{{\bf Anomaly non-cancellation in the static limit:} Net parity-odd current component as a function of the background field strength $E_x/E_{c}$. As a benchmark test, the analytic result (solid line) is compared to data points from lattice simulations (diamonds) in the corresponding static field limit.}
	\label{fig:mass}
\end{figure}

This phenomenon is illustrated in \Fig{fig:mass}, where we show the total parity-violating current over a wide range of field strengths. We note that $e E_x/(4\pi)$ is an upper limit for the net parity-odd current. Below we will see that our analytic results capture the earlier stages of the fully dynamical evolution, where they can also be used to benchmark numerical simulations, until back-action of the fermions on the electric field becomes relevant.

{\it Real-time lattice simulations with Wilson fermions.}
In the regime of large fields, real-time simulations in a Hamiltonian lattice formulation of QED$_{2+1}$ \cite{Kogut:1974ag} become feasible using 
classical-statistical reweighting techniques~\cite{Aarts:1998td,Berges:2010zv,Saffin:2011kc,Hebenstreit:2013qxa,Kasper:2014uaa,Buividovich:2015jfa,Gelfand:2016prm,Tanji:2016dka,Mueller:2016ven,Mueller:2016aao}.
Discretized on a spatially periodic lattice with $N^2$ sites and spacing $a$, the $O(a^3)$ improved Hamiltonian reads~\cite{Mueller:2016ven,Mueller:2016aao,Mace:2016shq,Spitz:2018eps} 
\begin{align}
\label{eq:LatticeHamiltonian}
& \hat{H}_{\{ m, r\}} = a^2 \sum_{n\in \mathbb{N}_0} C_n\Bigg( \sum_{\mathbf{x} \in \Lambda} \hat{\bar{\psi}}_{\mathbf{x}}^{\alpha} \Big(m + \frac{2nr}{a}\Big) \hat{\psi}_{\mathbf{x}}^{\alpha}\nonumber \\
&+  \sum_{\mathbf{x} \in \Lambda, \hat{i}} \frac{1}{2a} \left( \hat{\bar{\psi}}_{\mathbf{x}}^{\alpha}\left(i\gamma^{i}+n r\right)U_{\mathbf{x},n\cdot\hat{i}}\hat{\psi}_{\mathbf{x}+n\cdot\hat{i}}^{\alpha} + h.c.\right)\Bigg) .
\end{align}
Here $r$ is the Wilson parameter for the suppression of fermion doublers, $\Lambda = \{(x,y)|x,y\in0,\ldots,N-1\}$ the spatial lattice and $\hat{i}$ denotes the unit lattice vector in spatial direction $i = x,y$. The improvement coefficients are $C_0 = 1.5$, $C_1 = -0.3$, and $C_2 = -1/30$ with $C_{n>2} = 0$. As initial condition, we consider a free fermionic vacuum with homogeneous field $\mathbf{E}_0= (E_x,0)$, switched-on instantaneously at initial time $t=0$.
 
Both the mass and Wilson parameter break parity symmetry, and we extract the parity-odd current components from the total fermion electric current $j^i_{\text{total}}(t)$ by a linear combination of evolutions with different choices for the signs of $m$ and $r$~\cite{COSTE1989631}:
\begin{align}
\label{eq:current-definition}
j_{m}^{i}(t) & =  \frac{1}{2} \Big(j^i_{\text{total}} (t)_{\{+m,+r\}}-j^i_{\text{total}} (t)_{\{- m, + r\}}\Big) \; ,\nonumber \\
j_{\mathrm{an}}^{i}(t) & =  \frac{1}{2} \Big(j^i_{\text{total}} (t)_{\{+m,+r\}}-j^i_{\text{total}} (t)_{\{ + m,-r\}}\Big) \; .
\end{align}
Here, $j^i_{\text{total}} (t)_{\{\pm m,\pm r\}}$ refers to the expectation value of the lattice current operator as derived from the respective Hamiltonian $\hat{H}_{\{ \pm m,\pm r\}}$ given in Eq.~(\ref{eq:LatticeHamiltonian}). 

Specifically, we choose for the ``physical'' fermion $m,r >0$ to compute the current $j^i_{\text{total}} (t)_{\{+m,+r\}}$ and extract the corresponding gauge field evolution. The latter is then implemented as a (time-dependent) background field for evolutions of the ``test'' fermions having the other sign combinations required for $j^i_{\text{total}} (t)_{\{-m,+r\}}$ and $j^i_{\text{total}} (t)_{\{+m,-r\}}$. For the strong homogeneous fields employed in our study, this procedure provides a very efficent way to compute the parity-odd components~\footnote{For all simulations we employ spatially homogeneous electric field configurations. %and checked the stability of our results against adding the ``quantum-half'' fluctuations in the initial conditions.
}.

To benchmark our simulations, we first implement a static field approximation as correspondingly employed for our analytical results above. The simulation results for the net parity-odd current, which are extracted after a short period of time following the initial switch-on, for $e=0.1\sqrt{m}$, $N_f = 1$ and different values of the applied field strength are summarized in \Fig{fig:mass}~\footnote{We employ lattice parameters in the range of $ma = 0.025 - 0.1$ and $N^2 = 128 \times 128 - 256 \times 256$, depending on the actual field strength to ensure insensitivity to the lattice regularization.}. The data points agree to very good accuracy with the analytical predictions (\ref{eq:j-odd-Erf}) and (\ref{eq:j-odd}). 

\textit{Anomalous field rotation.}
Next we investigate the fully dynamical case including the mutual back-action of gauge and fermion fields. For field strengths beyond the Schwinger limit, the produced fermions and anti-fermions are accelerated to opposite directions, such that they counteract the initial electric field. The non-linear process leads to plasma oscillations, similar to what is found in 1+1 dimensional~\cite{Hebenstreit:2013qxa} and 3+1 dimensional lattice simulations~\cite{Kasper:2014uaa,Mueller:2016aao}. 

For QED$_{2+1}$ non-cancellation of the parity anomaly leads to a macroscopic non-zero transverse current. Because of the dynamical back-action the orientation of the electric field changes. As a consequence, the vectors describing the classical conduction current $j_{\mathrm{class}}$ and the electric field are no longer aligned, as illustrated in panel (c) of \Fig{fig:Schema}.
The interplay of dynamical parity-violation and classical conduction currents results in a non-linear rotation of the electric field vector, accompanied by a characteristic oscillation of its magnitude.
This is illustrated in \Fig{fig:Schema} (d) for initial field $\mathbf{E}_0 = (10 E_c,0)$  and $e=0.5\sqrt{m}$ \footnote{We employ $ma = 0.05$, $N^2=448\times448$ with fermions of the type $m,r>0$. Neither an increase or decrease of the UV-cutoff by a factor of two (at constant physical volume), nor a further increase of the lattice volume by $25\%$ lead to qualitative changes of our simulation result.}. Without loss of generality, we consider $N_f=10$ fermion flavors in order to accelerate the dynamics in view of limited computational resources \footnote{Here, all $N_f$ fermion flavors have the same mass, such that no current cancellations between different flavors with opposite mass signs occur.}.

\begin{figure}
	\includegraphics[width=0.49\textwidth]{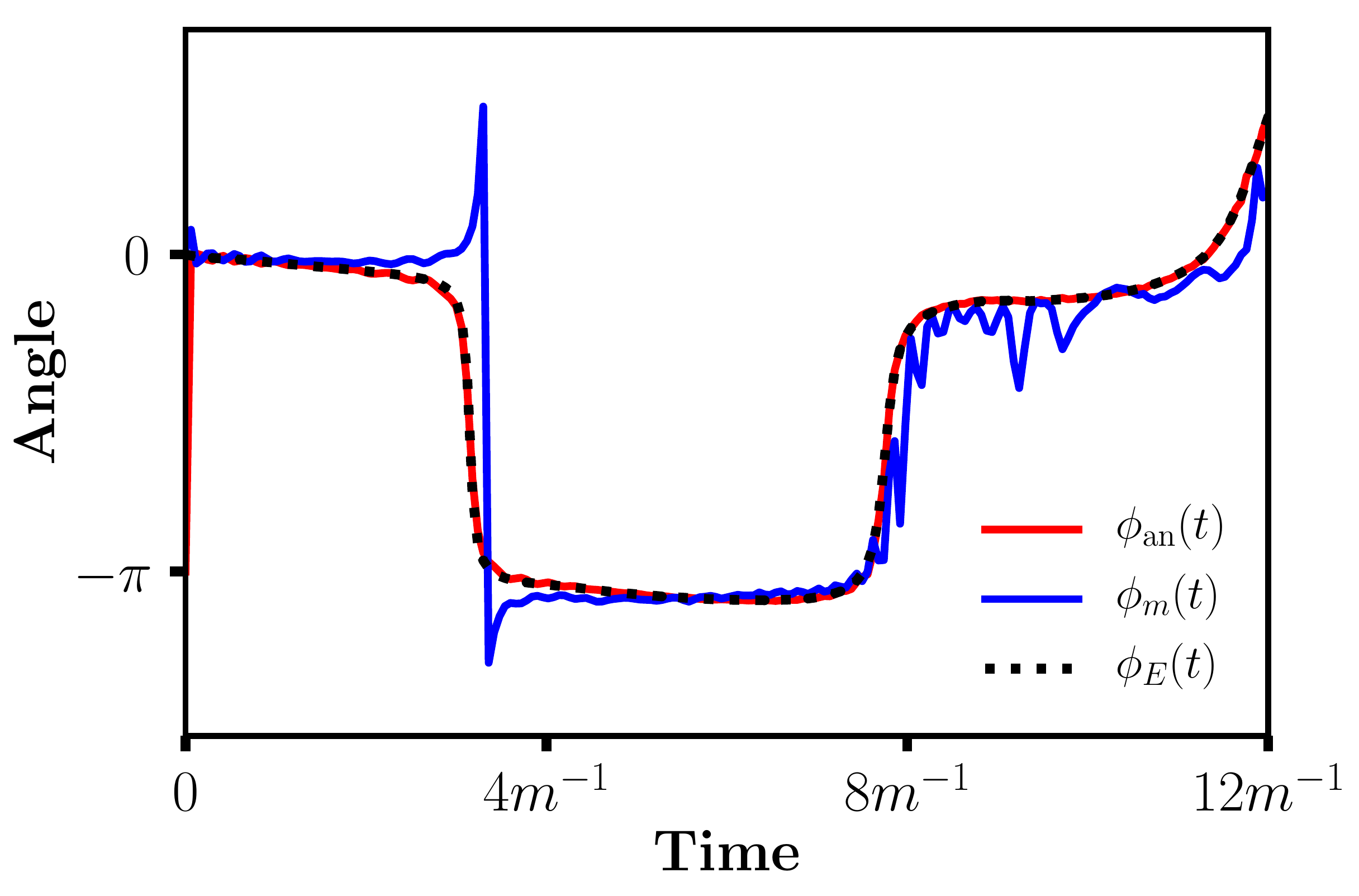}
	\caption{{\bf Anomalous field and currents rotations:} Time evolution of the orientation of the parity-odd currents from the anomaly ($\phi_{\text{an}}(t)$), the mass term ($\phi_m(t)$), and the electric field ($\phi_E(t)$) as introduced in \Fig{fig:Schema} (c), using lattice simulations with fully dynamical gauge field and fermions.}
	\label{fig:phase}
\end{figure}

Throughout the evolution parity-odd currents spend most of the time perpendicular to the electric field vector. This is quantified in \Fig{fig:phase}, which shows the orientation of the current and electric field vectors introduced in panel (c) of \Fig{fig:Schema}. This anomalous rotation of the electric field and currents represents a distinct signal of the violation of parity out of equilibrium.

\begin{figure}
	\centering
	\includegraphics[width=0.48\textwidth]{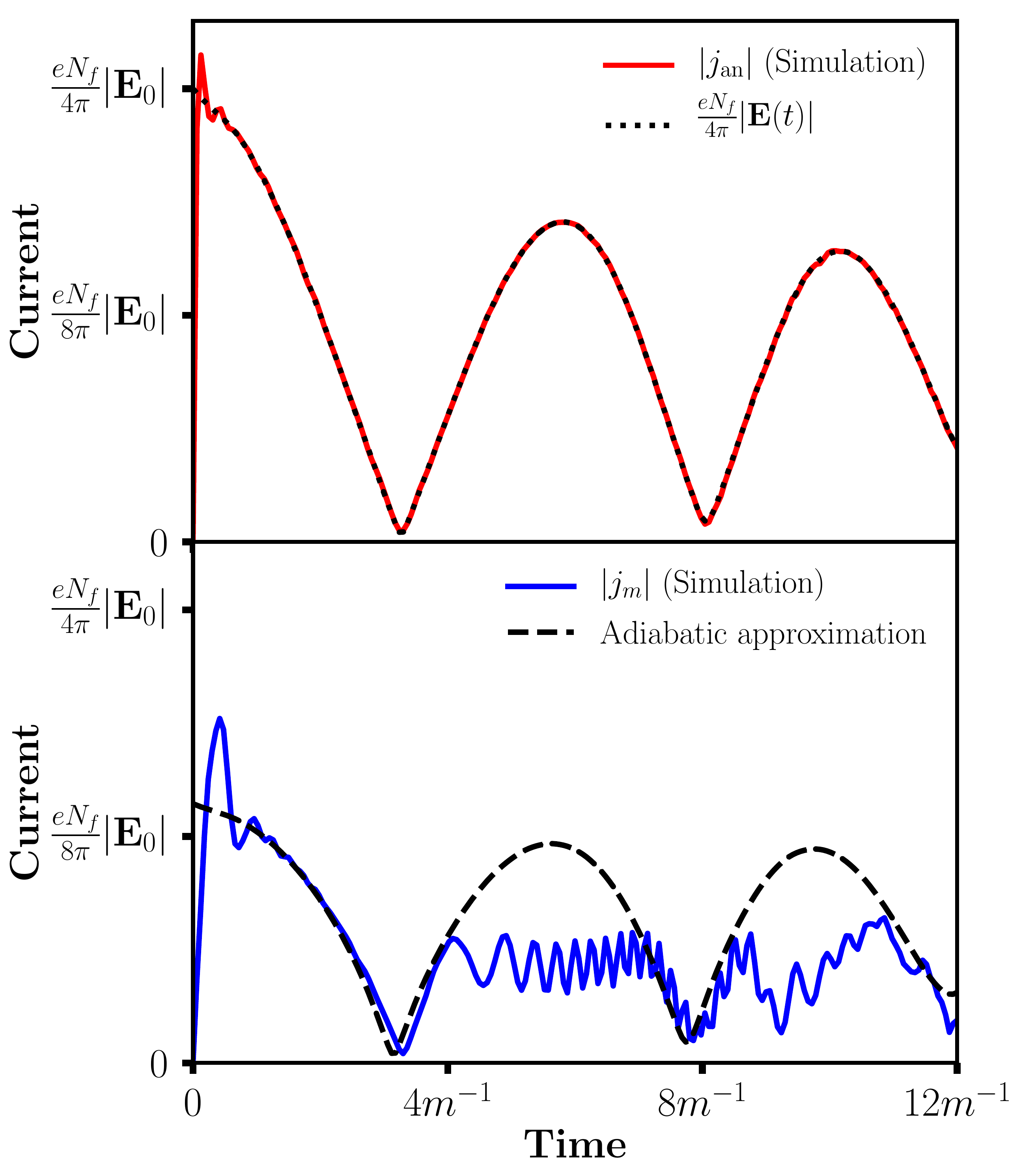}
	\caption{{\bf Anomaly non-cancellation for dynamical field:} (Top) Magnitude of the anomalous current $j_{\mathrm{an}}$ (red solid) for strong-field initial conditions, compared to the analytical prediction~(\ref{eq:j-odd}) evaluated for the dynamical electric field (dotted). (Bottom) The mass-induced parity-odd current $j_{m}$ (blue solid) is smaller in magnitude with  different time-dependence, such that anomaly cancellation does not occur. These fully dynamical simulation results are compared to Eq.~\eqref{eq:j-odd-Erf} with the static electric field replaced by $|\mathbf{E}(t)|$ (dashed), pointing out important corrections beyond the adiabatic approximation.}
	\label{fig:Anomaly}
\end{figure}

In \Fig{fig:Anomaly}, we display the absolute values of the spatial currents $|j_{\text{an}} (t)|$ (top panel) and $|j_\text{m} (t)|$ (bottom panel). Starting from zero at initial time, the currents build up quickly as a result of the applied field that is switched on at $t=0$. 
While $|j_{\text{an}} (t)|$ oscillates in phase with the total electric field strength, $|j_\text{m} (t)|$ is smaller in magnitude and develops a more complicated time-dependence, demonstrating that anomaly cancellation does not occur.

To gain analytical insights, we compare the fully dynamical simulation results to an adiabatic approximation, where we insert into our estimates \eqref{eq:j-odd-Erf} and (\ref{eq:j-odd}) the time-dependent $|\mathbf{E}(t)|$ obtained from the simulations. The upper panel of \Fig{fig:Anomaly} shows for $|j_{\text{an}}|$ a very good agreement of the adiabatic approximation with the full simulation data after a short initial period of time. The switching-on effect at $t\gtrsim 0$ is not captured by the analytical formulae, where the gauge field is initialized in the remote past. We checked explicitly that the agreement holds for a wide range of initial conditions for the gauge fields, confirming the anomaly far from equilibrium. 

In contrast, we observe differences between the full simulation results and the adiabatic approximation for the medium-modified current $j_\text{m} (t)$ once the fields change on time scales $\sim m^{-1}$, which we attribute to the effects of higher derivative terms neglected in \eqref{eq:j-odd-Erf}. For small gauge couplings employed for the results of \Fig{fig:Anomaly}, the evolution is comparably slow and the adiabatic approximation still captures important features of the $|j_\text{m} (t)|$ evolution for not too late times. The bottom graph of \Fig{fig:Anomaly} shows that the approximation agrees reasonably well with the lattice simulation data in a regime, where the electric field already begins to oscillate and its rotation is initiated. At later times, more rapid changes occur and significant deviations are seen. We find this phenomenon to coincide with particles approaching the zero mode in momentum space. In that case, there is interference with further pair creation in the low-momentum region and the adiabatic approximation is expected to break down~\footnote{In fact, similar effects have been observed in fermion distributions for time-dependent electric fields with sub-cycle structure \cite{Hebenstreit:2009km,Dumlu:2010ua}.}.

{\it Conclusions.} Our results provide an intriguing example where strong fields make genuine quantum phenomena visible on macroscopic scales, which are otherwise suppressed in near-vacuum conditions. Using ab initio simulations and analytic techniques, we established a non-cancellation of the parity anomaly for fields exceeding the Schwinger limit for pair production. Remarkably, we find that even aspects of the fully non-linear time-dependent processes can be approximately described with our adiabatic formulae. Together with the lattice simulation results these give unprecedented insights into the strong-field regime of QED$_{2+1}$ with important consequences for associated physical systems. 

Moreover, our study will be instrumental for further developments of quantum simulators for gauge theories with the help of synthetic materials based on atomic, molecular and optical physics engineering \cite{Wiese:2013uua,Zohar:2015hwa,Tagliacozzo:2012df,Tagliacozzo:2014bta,Haegeman:2014maa,Lamm:2019bik,BlattZoller,Zache:2018cqq,Zache:2018jbt,Klco:2018kyo,schweizer2019floquet}. There are strong international efforts towards quantum simulating QED in two spatial dimensions. Here the construction of magnetic plaquette terms on the lattice, which typically involve four-body processes, provides a particular challenge~\cite{PhysRevLett.118.070501,Zohar:2016iic,Zohar:2015hwa,Zohar:2013zla,Bender:2018rdp}. Our results indicate that the anomalous rotation of the electric field is practically not affected by the magnetic sector, such that first benchmark tests of genuinely $2+1$ dimensional quantum phenomena in QED with the next generation of quantum simulators may be realized with a drastically reduced experimental complexity.

{\it Acknowledgments.} We thank Philipp Hauke and Oscar Garcia-Montero for discussions and collaborations on related work.
This work is part of and supported by the DFG Collaborative Research Centre ``SFB 1225 (ISOQUANT)''.
The authors acknowledge support by the state of Baden-W\"urttemberg through bwHPC.
NM would like to thank ITP Heidelberg for their kind hospitality during the completion of this work. 
NM is supported by the U.S. Department of Energy, Office of Science, Office of Nuclear Physics, under contract No. DE- SC0012704 and by the Deutsche Forschungsgemeinschaft (DFG, German Research Foundation) - Project 404640738. 
\newline
\bibliographystyle{apsrev4-1} 
\bibliography{bibliography}

\begin{thebibliography}{61}%
\makeatletter
\providecommand \@ifxundefined [1]{%
 \@ifx{#1\undefined}
}%
\providecommand \@ifnum [1]{%
 \ifnum #1\expandafter \@firstoftwo
 \else \expandafter \@secondoftwo
 \fi
}%
\providecommand \@ifx [1]{%
 \ifx #1\expandafter \@firstoftwo
 \else \expandafter \@secondoftwo
 \fi
}%
\providecommand \natexlab [1]{#1}%
\providecommand \enquote  [1]{``#1''}%
\providecommand \bibnamefont  [1]{#1}%
\providecommand \bibfnamefont [1]{#1}%
\providecommand \citenamefont [1]{#1}%
\providecommand \href@noop [0]{\@secondoftwo}%
\providecommand \href [0]{\begingroup \@sanitize@url \@href}%
\providecommand \@href[1]{\@@startlink{#1}\@@href}%
\providecommand \@@href[1]{\endgroup#1\@@endlink}%
\providecommand \@sanitize@url [0]{\catcode `\\12\catcode `\$12\catcode
  `\&12\catcode `\#12\catcode `\^12\catcode `\_12\catcode `\%12\relax}%
\providecommand \@@startlink[1]{}%
\providecommand \@@endlink[0]{}%
\providecommand \url  [0]{\begingroup\@sanitize@url \@url }%
\providecommand \@url [1]{\endgroup\@href {#1}{\urlprefix }}%
\providecommand \urlprefix  [0]{URL }%
\providecommand \Eprint [0]{\href }%
\providecommand \doibase [0]{http://dx.doi.org/}%
\providecommand \selectlanguage [0]{\@gobble}%
\providecommand \bibinfo  [0]{\@secondoftwo}%
\providecommand \bibfield  [0]{\@secondoftwo}%
\providecommand \translation [1]{[#1]}%
\providecommand \BibitemOpen [0]{}%
\providecommand \bibitemStop [0]{}%
\providecommand \bibitemNoStop [0]{.\EOS\space}%
\providecommand \EOS [0]{\spacefactor3000\relax}%
\providecommand \BibitemShut  [1]{\csname bibitem#1\endcsname}%
\let\auto@bib@innerbib\@empty
%</preamble>
\bibitem [{\citenamefont {Li}\ \emph {et~al.}(2016)\citenamefont {Li},
  \citenamefont {Kharzeev}, \citenamefont {Zhang}, \citenamefont {Huang},
  \citenamefont {Pletikosi{\'c}}, \citenamefont {Fedorov}, \citenamefont
  {Zhong}, \citenamefont {Schneeloch}, \citenamefont {Gu},\ and\ \citenamefont
  {Valla}}]{Q.Lietal.}%
  \BibitemOpen
  \bibfield  {author} {\bibinfo {author} {\bibfnamefont {Q.}~\bibnamefont
  {Li}}, \bibinfo {author} {\bibfnamefont {D.~E.}\ \bibnamefont {Kharzeev}},
  \bibinfo {author} {\bibfnamefont {C.}~\bibnamefont {Zhang}}, \bibinfo
  {author} {\bibfnamefont {Y.}~\bibnamefont {Huang}}, \bibinfo {author}
  {\bibfnamefont {I.}~\bibnamefont {Pletikosi{\'c}}}, \bibinfo {author}
  {\bibfnamefont {A.~V.}\ \bibnamefont {Fedorov}}, \bibinfo {author}
  {\bibfnamefont {R.~D.}\ \bibnamefont {Zhong}}, \bibinfo {author}
  {\bibfnamefont {J.~A.}\ \bibnamefont {Schneeloch}}, \bibinfo {author}
  {\bibfnamefont {G.~D.}\ \bibnamefont {Gu}}, \ and\ \bibinfo {author}
  {\bibfnamefont {T.}~\bibnamefont {Valla}},\ }\href
  {https://doi.org/10.1038/nphys3648} {\bibfield  {journal} {\bibinfo
  {journal} {Nature Physics}\ }\textbf {\bibinfo {volume} {12}},\ \bibinfo
  {pages} {550 EP } (\bibinfo {year} {2016})}\BibitemShut {NoStop}%
\bibitem [{\citenamefont {Xiong}\ \emph {et~al.}(2015)\citenamefont {Xiong},
  \citenamefont {Kushwaha}, \citenamefont {Liang}, \citenamefont {Krizan},
  \citenamefont {Hirschberger}, \citenamefont {Wang}, \citenamefont {Cava},\
  and\ \citenamefont {Ong}}]{J.Xiongetal.}%
  \BibitemOpen
  \bibfield  {author} {\bibinfo {author} {\bibfnamefont {J.}~\bibnamefont
  {Xiong}}, \bibinfo {author} {\bibfnamefont {S.~K.}\ \bibnamefont {Kushwaha}},
  \bibinfo {author} {\bibfnamefont {T.}~\bibnamefont {Liang}}, \bibinfo
  {author} {\bibfnamefont {J.~W.}\ \bibnamefont {Krizan}}, \bibinfo {author}
  {\bibfnamefont {M.}~\bibnamefont {Hirschberger}}, \bibinfo {author}
  {\bibfnamefont {W.}~\bibnamefont {Wang}}, \bibinfo {author} {\bibfnamefont
  {R.~J.}\ \bibnamefont {Cava}}, \ and\ \bibinfo {author} {\bibfnamefont
  {N.~P.}\ \bibnamefont {Ong}},\ }\href {\doibase 10.1126/science.aac6089}
  {\bibfield  {journal} {\bibinfo  {journal} {Science}\ }\textbf {\bibinfo
  {volume} {350}},\ \bibinfo {pages} {413} (\bibinfo {year} {2015})},\ \Eprint
  {http://arxiv.org/abs/http://science.sciencemag.org/content/350/6259/413.full.pdf}
  {http://science.sciencemag.org/content/350/6259/413.full.pdf} \BibitemShut
  {NoStop}%
\bibitem [{\citenamefont {Hosur}\ and\ \citenamefont
  {Qi}(2013)}]{Hosur:2013kxa}%
  \BibitemOpen
  \bibfield  {author} {\bibinfo {author} {\bibfnamefont {P.}~\bibnamefont
  {Hosur}}\ and\ \bibinfo {author} {\bibfnamefont {X.}~\bibnamefont {Qi}},\
  }\href {\doibase 10.1016/j.crhy.2013.10.010} {\bibfield  {journal} {\bibinfo
  {journal} {Comptes Rendus Physique}\ }\textbf {\bibinfo {volume} {14}},\
  \bibinfo {pages} {857} (\bibinfo {year} {2013})},\ \Eprint
  {http://arxiv.org/abs/1309.4464} {arXiv:1309.4464 [cond-mat.str-el]}
  \BibitemShut {NoStop}%
%%CITATION = ARXIV:1309.4464;%%
\bibitem [{\citenamefont {Gorbar}\ \emph {et~al.}(2018)\citenamefont {Gorbar},
  \citenamefont {Miransky}, \citenamefont {Shovkovy},\ and\ \citenamefont
  {Sukhachov}}]{Gorbar:2017lnp}%
  \BibitemOpen
  \bibfield  {author} {\bibinfo {author} {\bibfnamefont {E.~V.}\ \bibnamefont
  {Gorbar}}, \bibinfo {author} {\bibfnamefont {V.~A.}\ \bibnamefont
  {Miransky}}, \bibinfo {author} {\bibfnamefont {I.~A.}\ \bibnamefont
  {Shovkovy}}, \ and\ \bibinfo {author} {\bibfnamefont {P.~O.}\ \bibnamefont
  {Sukhachov}},\ }\href {\doibase 10.1063/1.5037551} {\bibfield  {journal}
  {\bibinfo  {journal} {Low Temp. Phys.}\ }\textbf {\bibinfo {volume} {44}},\
  \bibinfo {pages} {487} (\bibinfo {year} {2018})},\ \bibinfo {note} {[Fiz.
  Nizk. Temp.44,635(2017)]},\ \Eprint {http://arxiv.org/abs/1712.08947}
  {arXiv:1712.08947 [cond-mat.mes-hall]} \BibitemShut {NoStop}%
%%CITATION = ARXIV:1712.08947;%%
\bibitem [{\citenamefont {Kharzeev}\ \emph {et~al.}(2008)\citenamefont
  {Kharzeev}, \citenamefont {McLerran},\ and\ \citenamefont
  {Warringa}}]{Kharzeev:2007jp}%
  \BibitemOpen
  \bibfield  {author} {\bibinfo {author} {\bibfnamefont {D.~E.}\ \bibnamefont
  {Kharzeev}}, \bibinfo {author} {\bibfnamefont {L.~D.}\ \bibnamefont
  {McLerran}}, \ and\ \bibinfo {author} {\bibfnamefont {H.~J.}\ \bibnamefont
  {Warringa}},\ }\href {\doibase 10.1016/j.nuclphysa.2008.02.298} {\bibfield
  {journal} {\bibinfo  {journal} {Nucl. Phys.}\ }\textbf {\bibinfo {volume}
  {A803}},\ \bibinfo {pages} {227} (\bibinfo {year} {2008})},\ \Eprint
  {http://arxiv.org/abs/0711.0950} {arXiv:0711.0950 [hep-ph]} \BibitemShut
  {NoStop}%
%%CITATION = ARXIV:0711.0950;%%
\bibitem [{\citenamefont {Fukushima}\ \emph {et~al.}(2008)\citenamefont
  {Fukushima}, \citenamefont {Kharzeev},\ and\ \citenamefont
  {Warringa}}]{Fukushima:2008xe}%
  \BibitemOpen
  \bibfield  {author} {\bibinfo {author} {\bibfnamefont {K.}~\bibnamefont
  {Fukushima}}, \bibinfo {author} {\bibfnamefont {D.~E.}\ \bibnamefont
  {Kharzeev}}, \ and\ \bibinfo {author} {\bibfnamefont {H.~J.}\ \bibnamefont
  {Warringa}},\ }\href {\doibase 10.1103/PhysRevD.78.074033} {\bibfield
  {journal} {\bibinfo  {journal} {Phys. Rev.}\ }\textbf {\bibinfo {volume}
  {D78}},\ \bibinfo {pages} {074033} (\bibinfo {year} {2008})},\ \Eprint
  {http://arxiv.org/abs/0808.3382} {arXiv:0808.3382 [hep-ph]} \BibitemShut
  {NoStop}%
%%CITATION = ARXIV:0808.3382;%%
\bibitem [{\citenamefont {Kharzeev}\ \emph {et~al.}(2016)\citenamefont
  {Kharzeev}, \citenamefont {Liao}, \citenamefont {Voloshin},\ and\
  \citenamefont {Wang}}]{Kharzeev:2015znc}%
  \BibitemOpen
  \bibfield  {author} {\bibinfo {author} {\bibfnamefont {D.~E.}\ \bibnamefont
  {Kharzeev}}, \bibinfo {author} {\bibfnamefont {J.}~\bibnamefont {Liao}},
  \bibinfo {author} {\bibfnamefont {S.~A.}\ \bibnamefont {Voloshin}}, \ and\
  \bibinfo {author} {\bibfnamefont {G.}~\bibnamefont {Wang}},\ }\href {\doibase
  10.1016/j.ppnp.2016.01.001} {\bibfield  {journal} {\bibinfo  {journal} {Prog.
  Part. Nucl. Phys.}\ }\textbf {\bibinfo {volume} {88}},\ \bibinfo {pages} {1}
  (\bibinfo {year} {2016})},\ \Eprint {http://arxiv.org/abs/1511.04050}
  {arXiv:1511.04050 [hep-ph]} \BibitemShut {NoStop}%
%%CITATION = ARXIV:1511.04050;%%
\bibitem [{\citenamefont {Koch}\ \emph {et~al.}(2017)\citenamefont {Koch},
  \citenamefont {Schlichting}, \citenamefont {Skokov}, \citenamefont
  {Sorensen}, \citenamefont {Thomas}, \citenamefont {Voloshin}, \citenamefont
  {Wang},\ and\ \citenamefont {Yee}}]{Skokov:2016yrj}%
  \BibitemOpen
  \bibfield  {author} {\bibinfo {author} {\bibfnamefont {V.}~\bibnamefont
  {Koch}}, \bibinfo {author} {\bibfnamefont {S.}~\bibnamefont {Schlichting}},
  \bibinfo {author} {\bibfnamefont {V.}~\bibnamefont {Skokov}}, \bibinfo
  {author} {\bibfnamefont {P.}~\bibnamefont {Sorensen}}, \bibinfo {author}
  {\bibfnamefont {J.}~\bibnamefont {Thomas}}, \bibinfo {author} {\bibfnamefont
  {S.}~\bibnamefont {Voloshin}}, \bibinfo {author} {\bibfnamefont
  {G.}~\bibnamefont {Wang}}, \ and\ \bibinfo {author} {\bibfnamefont {H.-U.}\
  \bibnamefont {Yee}},\ }\href {\doibase 10.1088/1674-1137/41/7/072001}
  {\bibfield  {journal} {\bibinfo  {journal} {Chin. Phys.}\ }\textbf {\bibinfo
  {volume} {C41}},\ \bibinfo {pages} {072001} (\bibinfo {year} {2017})},\
  \Eprint {http://arxiv.org/abs/1608.00982} {arXiv:1608.00982 [nucl-th]}
  \BibitemShut {NoStop}%
%%CITATION = ARXIV:1608.00982;%%
\bibitem [{\citenamefont {Blaum}(2006)}]{BLAUM20061}%
  \BibitemOpen
  \bibfield  {author} {\bibinfo {author} {\bibfnamefont {K.}~\bibnamefont
  {Blaum}},\ }\href {\doibase https://doi.org/10.1016/j.physrep.2005.10.011}
  {\bibfield  {journal} {\bibinfo  {journal} {Physics Reports}\ }\textbf
  {\bibinfo {volume} {425}},\ \bibinfo {pages} {1 } (\bibinfo {year}
  {2006})}\BibitemShut {NoStop}%
\bibitem [{\citenamefont {Sturm}\ \emph {et~al.}(2014)\citenamefont {Sturm},
  \citenamefont {K{\"o}hler}, \citenamefont {Zatorski}, \citenamefont {Wagner},
  \citenamefont {Harman}, \citenamefont {Werth}, \citenamefont {Quint},
  \citenamefont {Keitel},\ and\ \citenamefont {Blaum}}]{BLAUMNature}%
  \BibitemOpen
  \bibfield  {author} {\bibinfo {author} {\bibfnamefont {S.}~\bibnamefont
  {Sturm}}, \bibinfo {author} {\bibfnamefont {F.}~\bibnamefont {K{\"o}hler}},
  \bibinfo {author} {\bibfnamefont {J.}~\bibnamefont {Zatorski}}, \bibinfo
  {author} {\bibfnamefont {A.}~\bibnamefont {Wagner}}, \bibinfo {author}
  {\bibfnamefont {Z.}~\bibnamefont {Harman}}, \bibinfo {author} {\bibfnamefont
  {G.}~\bibnamefont {Werth}}, \bibinfo {author} {\bibfnamefont
  {W.}~\bibnamefont {Quint}}, \bibinfo {author} {\bibfnamefont {C.~H.}\
  \bibnamefont {Keitel}}, \ and\ \bibinfo {author} {\bibfnamefont
  {K.}~\bibnamefont {Blaum}},\ }\href {https://doi.org/10.1038/nature13026}
  {\bibfield  {journal} {\bibinfo  {journal} {Nature}\ }\textbf {\bibinfo
  {volume} {506}},\ \bibinfo {pages} {467 EP } (\bibinfo {year}
  {2014})}\BibitemShut {NoStop}%
\bibitem [{\citenamefont {Di~Piazza}\ \emph {et~al.}(2012)\citenamefont
  {Di~Piazza}, \citenamefont {Muller}, \citenamefont {Hatsagortsyan},\ and\
  \citenamefont {Keitel}}]{DiPiazza:2011tq}%
  \BibitemOpen
  \bibfield  {author} {\bibinfo {author} {\bibfnamefont {A.}~\bibnamefont
  {Di~Piazza}}, \bibinfo {author} {\bibfnamefont {C.}~\bibnamefont {Muller}},
  \bibinfo {author} {\bibfnamefont {K.~Z.}\ \bibnamefont {Hatsagortsyan}}, \
  and\ \bibinfo {author} {\bibfnamefont {C.~H.}\ \bibnamefont {Keitel}},\
  }\href {\doibase 10.1103/RevModPhys.84.1177} {\bibfield  {journal} {\bibinfo
  {journal} {Rev. Mod. Phys.}\ }\textbf {\bibinfo {volume} {84}},\ \bibinfo
  {pages} {1177} (\bibinfo {year} {2012})},\ \Eprint
  {http://arxiv.org/abs/1111.3886} {arXiv:1111.3886 [hep-ph]} \BibitemShut
  {NoStop}%
%%CITATION = ARXIV:1111.3886;%%
\bibitem [{\citenamefont {Schwinger}(1951)}]{PhysRev.82.664}%
  \BibitemOpen
  \bibfield  {author} {\bibinfo {author} {\bibfnamefont {J.}~\bibnamefont
  {Schwinger}},\ }\href {\doibase 10.1103/PhysRev.82.664} {\bibfield  {journal}
  {\bibinfo  {journal} {Phys. Rev.}\ }\textbf {\bibinfo {volume} {82}},\
  \bibinfo {pages} {664} (\bibinfo {year} {1951})}\BibitemShut {NoStop}%
\bibitem [{\citenamefont {Wiese}(2013)}]{Wiese:2013uua}%
  \BibitemOpen
  \bibfield  {author} {\bibinfo {author} {\bibfnamefont {U.-J.}\ \bibnamefont
  {Wiese}},\ }\href {\doibase 10.1002/andp.201300104} {\bibfield  {journal}
  {\bibinfo  {journal} {Annalen Phys.}\ }\textbf {\bibinfo {volume} {525}},\
  \bibinfo {pages} {777} (\bibinfo {year} {2013})},\ \Eprint
  {http://arxiv.org/abs/1305.1602} {arXiv:1305.1602 [quant-ph]} \BibitemShut
  {NoStop}%
%%CITATION = ARXIV:1305.1602;%%
\bibitem [{\citenamefont {Zohar}\ \emph {et~al.}(2016)\citenamefont {Zohar},
  \citenamefont {Cirac},\ and\ \citenamefont {Reznik}}]{Zohar:2015hwa}%
  \BibitemOpen
  \bibfield  {author} {\bibinfo {author} {\bibfnamefont {E.}~\bibnamefont
  {Zohar}}, \bibinfo {author} {\bibfnamefont {J.~I.}\ \bibnamefont {Cirac}}, \
  and\ \bibinfo {author} {\bibfnamefont {B.}~\bibnamefont {Reznik}},\ }\href
  {\doibase 10.1088/0034-4885/79/1/014401} {\bibfield  {journal} {\bibinfo
  {journal} {Rept. Prog. Phys.}\ }\textbf {\bibinfo {volume} {79}},\ \bibinfo
  {pages} {014401} (\bibinfo {year} {2016})},\ \Eprint
  {http://arxiv.org/abs/1503.02312} {arXiv:1503.02312 [quant-ph]} \BibitemShut
  {NoStop}%
%%CITATION = ARXIV:1503.02312;%%
\bibitem [{\citenamefont {Cirac}\ and\ \citenamefont
  {Zoller}(2012)}]{CiracZoller}%
  \BibitemOpen
  \bibfield  {author} {\bibinfo {author} {\bibfnamefont {J.~I.}\ \bibnamefont
  {Cirac}}\ and\ \bibinfo {author} {\bibfnamefont {P.}~\bibnamefont {Zoller}},\
  }\href {https://doi.org/10.1038/nphys2275} {\bibfield  {journal} {\bibinfo
  {journal} {Nature Physics}\ }\textbf {\bibinfo {volume} {8}},\ \bibinfo
  {pages} {264 EP } (\bibinfo {year} {2012})}\BibitemShut {NoStop}%
\bibitem [{\citenamefont {Hauke}\ \emph {et~al.}(2012)\citenamefont {Hauke},
  \citenamefont {Cucchietti}, \citenamefont {Tagliacozzo}, \citenamefont
  {Deutsch},\ and\ \citenamefont {Lewenstein}}]{Hauke_2012}%
  \BibitemOpen
  \bibfield  {author} {\bibinfo {author} {\bibfnamefont {P.}~\bibnamefont
  {Hauke}}, \bibinfo {author} {\bibfnamefont {F.~M.}\ \bibnamefont
  {Cucchietti}}, \bibinfo {author} {\bibfnamefont {L.}~\bibnamefont
  {Tagliacozzo}}, \bibinfo {author} {\bibfnamefont {I.}~\bibnamefont
  {Deutsch}}, \ and\ \bibinfo {author} {\bibfnamefont {M.}~\bibnamefont
  {Lewenstein}},\ }\href {\doibase 10.1088/0034-4885/75/8/082401} {\bibfield
  {journal} {\bibinfo  {journal} {Reports on Progress in Physics}\ }\textbf
  {\bibinfo {volume} {75}},\ \bibinfo {pages} {082401} (\bibinfo {year}
  {2012})}\BibitemShut {NoStop}%
\bibitem [{\citenamefont {Tagliacozzo}\ \emph {et~al.}(2013)\citenamefont
  {Tagliacozzo}, \citenamefont {Celi}, \citenamefont {Orland},\ and\
  \citenamefont {Lewenstein}}]{Tagliacozzo:2012df}%
  \BibitemOpen
  \bibfield  {author} {\bibinfo {author} {\bibfnamefont {L.}~\bibnamefont
  {Tagliacozzo}}, \bibinfo {author} {\bibfnamefont {A.}~\bibnamefont {Celi}},
  \bibinfo {author} {\bibfnamefont {P.}~\bibnamefont {Orland}}, \ and\ \bibinfo
  {author} {\bibfnamefont {M.}~\bibnamefont {Lewenstein}},\ }\href {\doibase
  10.1038/ncomms3615} {\bibfield  {journal} {\bibinfo  {journal} {Nature
  Commun.}\ }\textbf {\bibinfo {volume} {4}},\ \bibinfo {pages} {2615}
  (\bibinfo {year} {2013})},\ \Eprint {http://arxiv.org/abs/1211.2704}
  {arXiv:1211.2704 [cond-mat.quant-gas]} \BibitemShut {NoStop}%
%%CITATION = ARXIV:1211.2704;%%
\bibitem [{\citenamefont {Tagliacozzo}\ \emph {et~al.}(2014)\citenamefont
  {Tagliacozzo}, \citenamefont {Celi},\ and\ \citenamefont
  {Lewenstein}}]{Tagliacozzo:2014bta}%
  \BibitemOpen
  \bibfield  {author} {\bibinfo {author} {\bibfnamefont {L.}~\bibnamefont
  {Tagliacozzo}}, \bibinfo {author} {\bibfnamefont {A.}~\bibnamefont {Celi}}, \
  and\ \bibinfo {author} {\bibfnamefont {M.}~\bibnamefont {Lewenstein}},\
  }\href {\doibase 10.1103/PhysRevX.4.041024} {\bibfield  {journal} {\bibinfo
  {journal} {Phys. Rev.}\ }\textbf {\bibinfo {volume} {X4}},\ \bibinfo {pages}
  {041024} (\bibinfo {year} {2014})},\ \Eprint {http://arxiv.org/abs/1405.4811}
  {arXiv:1405.4811 [cond-mat.str-el]} \BibitemShut {NoStop}%
%%CITATION = ARXIV:1405.4811;%%
\bibitem [{\citenamefont {Haegeman}\ \emph {et~al.}(2015)\citenamefont
  {Haegeman}, \citenamefont {Van~Acoleyen}, \citenamefont {Schuch},
  \citenamefont {Cirac},\ and\ \citenamefont {Verstraete}}]{Haegeman:2014maa}%
  \BibitemOpen
  \bibfield  {author} {\bibinfo {author} {\bibfnamefont {J.}~\bibnamefont
  {Haegeman}}, \bibinfo {author} {\bibfnamefont {K.}~\bibnamefont
  {Van~Acoleyen}}, \bibinfo {author} {\bibfnamefont {N.}~\bibnamefont
  {Schuch}}, \bibinfo {author} {\bibfnamefont {J.~I.}\ \bibnamefont {Cirac}}, \
  and\ \bibinfo {author} {\bibfnamefont {F.}~\bibnamefont {Verstraete}},\
  }\href {\doibase 10.1103/PhysRevX.5.011024} {\bibfield  {journal} {\bibinfo
  {journal} {Phys. Rev.}\ }\textbf {\bibinfo {volume} {X5}},\ \bibinfo {pages}
  {011024} (\bibinfo {year} {2015})},\ \Eprint {http://arxiv.org/abs/1407.1025}
  {arXiv:1407.1025 [quant-ph]} \BibitemShut {NoStop}%
%%CITATION = ARXIV:1407.1025;%%
\bibitem [{\citenamefont {Lamm}\ \emph {et~al.}(2019)\citenamefont {Lamm},
  \citenamefont {Lawrence},\ and\ \citenamefont {Yamauchi}}]{Lamm:2019bik}%
  \BibitemOpen
  \bibfield  {author} {\bibinfo {author} {\bibfnamefont {H.}~\bibnamefont
  {Lamm}}, \bibinfo {author} {\bibfnamefont {S.}~\bibnamefont {Lawrence}}, \
  and\ \bibinfo {author} {\bibfnamefont {Y.}~\bibnamefont {Yamauchi}} (\bibinfo
  {collaboration} {NuQS}),\ }\href@noop {} {\  (\bibinfo {year} {2019})},\
  \Eprint {http://arxiv.org/abs/1903.08807} {arXiv:1903.08807 [hep-lat]}
  \BibitemShut {NoStop}%
%%CITATION = ARXIV:1903.08807;%%
\bibitem [{\citenamefont {Martinez}\ \emph {et~al.}(2016)\citenamefont
  {Martinez}, \citenamefont {Muschik}, \citenamefont {Schindler}, \citenamefont
  {Nigg}, \citenamefont {Erhard}, \citenamefont {Heyl}, \citenamefont {Hauke},
  \citenamefont {Dalmonte}, \citenamefont {Monz}, \citenamefont {Zoller},\ and\
  \citenamefont {Blatt}}]{BlattZoller}%
  \BibitemOpen
  \bibfield  {author} {\bibinfo {author} {\bibfnamefont {E.~A.}\ \bibnamefont
  {Martinez}}, \bibinfo {author} {\bibfnamefont {C.~A.}\ \bibnamefont
  {Muschik}}, \bibinfo {author} {\bibfnamefont {P.}~\bibnamefont {Schindler}},
  \bibinfo {author} {\bibfnamefont {D.}~\bibnamefont {Nigg}}, \bibinfo {author}
  {\bibfnamefont {A.}~\bibnamefont {Erhard}}, \bibinfo {author} {\bibfnamefont
  {M.}~\bibnamefont {Heyl}}, \bibinfo {author} {\bibfnamefont {P.}~\bibnamefont
  {Hauke}}, \bibinfo {author} {\bibfnamefont {M.}~\bibnamefont {Dalmonte}},
  \bibinfo {author} {\bibfnamefont {T.}~\bibnamefont {Monz}}, \bibinfo {author}
  {\bibfnamefont {P.}~\bibnamefont {Zoller}}, \ and\ \bibinfo {author}
  {\bibfnamefont {R.}~\bibnamefont {Blatt}},\ }\href
  {https://doi.org/10.1038/nature18318} {\bibfield  {journal} {\bibinfo
  {journal} {Nature}\ }\textbf {\bibinfo {volume} {534}},\ \bibinfo {pages}
  {516 EP } (\bibinfo {year} {2016})}\BibitemShut {NoStop}%
\bibitem [{\citenamefont {Zohar}\ \emph
  {et~al.}(2017{\natexlab{a}})\citenamefont {Zohar}, \citenamefont {Farace},
  \citenamefont {Reznik},\ and\ \citenamefont
  {Cirac}}]{PhysRevLett.118.070501}%
  \BibitemOpen
  \bibfield  {author} {\bibinfo {author} {\bibfnamefont {E.}~\bibnamefont
  {Zohar}}, \bibinfo {author} {\bibfnamefont {A.}~\bibnamefont {Farace}},
  \bibinfo {author} {\bibfnamefont {B.}~\bibnamefont {Reznik}}, \ and\ \bibinfo
  {author} {\bibfnamefont {J.~I.}\ \bibnamefont {Cirac}},\ }\href {\doibase
  10.1103/PhysRevLett.118.070501} {\bibfield  {journal} {\bibinfo  {journal}
  {Phys. Rev. Lett.}\ }\textbf {\bibinfo {volume} {118}},\ \bibinfo {pages}
  {070501} (\bibinfo {year} {2017}{\natexlab{a}})}\BibitemShut {NoStop}%
\bibitem [{\citenamefont {Zohar}\ \emph
  {et~al.}(2017{\natexlab{b}})\citenamefont {Zohar}, \citenamefont {Farace},
  \citenamefont {Reznik},\ and\ \citenamefont {Cirac}}]{Zohar:2016iic}%
  \BibitemOpen
  \bibfield  {author} {\bibinfo {author} {\bibfnamefont {E.}~\bibnamefont
  {Zohar}}, \bibinfo {author} {\bibfnamefont {A.}~\bibnamefont {Farace}},
  \bibinfo {author} {\bibfnamefont {B.}~\bibnamefont {Reznik}}, \ and\ \bibinfo
  {author} {\bibfnamefont {J.~I.}\ \bibnamefont {Cirac}},\ }\href {\doibase
  10.1103/PhysRevA.95.023604} {\bibfield  {journal} {\bibinfo  {journal} {Phys.
  Rev.}\ }\textbf {\bibinfo {volume} {A95}},\ \bibinfo {pages} {023604}
  (\bibinfo {year} {2017}{\natexlab{b}})},\ \Eprint
  {http://arxiv.org/abs/1607.08121} {arXiv:1607.08121 [quant-ph]} \BibitemShut
  {NoStop}%
%%CITATION = ARXIV:1607.08121;%%
\bibitem [{\citenamefont {Pichler}\ \emph {et~al.}(2016)\citenamefont
  {Pichler}, \citenamefont {Dalmonte}, \citenamefont {Rico}, \citenamefont
  {Zoller},\ and\ \citenamefont {Montangero}}]{Pichler:2015yqa}%
  \BibitemOpen
  \bibfield  {author} {\bibinfo {author} {\bibfnamefont {T.}~\bibnamefont
  {Pichler}}, \bibinfo {author} {\bibfnamefont {M.}~\bibnamefont {Dalmonte}},
  \bibinfo {author} {\bibfnamefont {E.}~\bibnamefont {Rico}}, \bibinfo {author}
  {\bibfnamefont {P.}~\bibnamefont {Zoller}}, \ and\ \bibinfo {author}
  {\bibfnamefont {S.}~\bibnamefont {Montangero}},\ }\href {\doibase
  10.1103/PhysRevX.6.011023} {\bibfield  {journal} {\bibinfo  {journal} {Phys.
  Rev.}\ }\textbf {\bibinfo {volume} {X6}},\ \bibinfo {pages} {011023}
  (\bibinfo {year} {2016})},\ \Eprint {http://arxiv.org/abs/1505.04440}
  {arXiv:1505.04440 [cond-mat.quant-gas]} \BibitemShut {NoStop}%
%%CITATION = ARXIV:1505.04440;%%
\bibitem [{\citenamefont {Hebenstreit}\ \emph
  {et~al.}(2013{\natexlab{a}})\citenamefont {Hebenstreit}, \citenamefont
  {Berges},\ and\ \citenamefont {Gelfand}}]{Hebenstreit:2013baa}%
  \BibitemOpen
  \bibfield  {author} {\bibinfo {author} {\bibfnamefont {F.}~\bibnamefont
  {Hebenstreit}}, \bibinfo {author} {\bibfnamefont {J.}~\bibnamefont {Berges}},
  \ and\ \bibinfo {author} {\bibfnamefont {D.}~\bibnamefont {Gelfand}},\ }\href
  {\doibase 10.1103/PhysRevLett.111.201601} {\bibfield  {journal} {\bibinfo
  {journal} {Phys. Rev. Lett.}\ }\textbf {\bibinfo {volume} {111}},\ \bibinfo
  {pages} {201601} (\bibinfo {year} {2013}{\natexlab{a}})},\ \Eprint
  {http://arxiv.org/abs/1307.4619} {arXiv:1307.4619 [hep-ph]} \BibitemShut
  {NoStop}%
%%CITATION = ARXIV:1307.4619;%%
\bibitem [{\citenamefont {Zache}\ \emph {et~al.}(2018)\citenamefont {Zache},
  \citenamefont {Hebenstreit}, \citenamefont {Jendrzejewski}, \citenamefont
  {Oberthaler}, \citenamefont {Berges},\ and\ \citenamefont
  {Hauke}}]{Zache:2018jbt}%
  \BibitemOpen
  \bibfield  {author} {\bibinfo {author} {\bibfnamefont {T.~V.}\ \bibnamefont
  {Zache}}, \bibinfo {author} {\bibfnamefont {F.}~\bibnamefont {Hebenstreit}},
  \bibinfo {author} {\bibfnamefont {F.}~\bibnamefont {Jendrzejewski}}, \bibinfo
  {author} {\bibfnamefont {M.~K.}\ \bibnamefont {Oberthaler}}, \bibinfo
  {author} {\bibfnamefont {J.}~\bibnamefont {Berges}}, \ and\ \bibinfo {author}
  {\bibfnamefont {P.}~\bibnamefont {Hauke}},\ }\href {\doibase
  10.1088/2058-9565/aac33b} {\bibfield  {journal} {\bibinfo  {journal} {Sci.
  Technol.}\ }\textbf {\bibinfo {volume} {3}},\ \bibinfo {pages} {034010}
  (\bibinfo {year} {2018})},\ \Eprint {http://arxiv.org/abs/1802.06704}
  {arXiv:1802.06704 [cond-mat.quant-gas]} \BibitemShut {NoStop}%
%%CITATION = ARXIV:1802.06704;%%
\bibitem [{\citenamefont {Coste}\ and\ \citenamefont
  {Lüscher}(1989)}]{COSTE1989631}%
  \BibitemOpen
  \bibfield  {author} {\bibinfo {author} {\bibfnamefont {A.}~\bibnamefont
  {Coste}}\ and\ \bibinfo {author} {\bibfnamefont {M.}~\bibnamefont
  {Lüscher}},\ }\href {\doibase https://doi.org/10.1016/0550-3213(89)90127-2}
  {\bibfield  {journal} {\bibinfo  {journal} {Nuclear Physics B}\ }\textbf
  {\bibinfo {volume} {323}},\ \bibinfo {pages} {631 } (\bibinfo {year}
  {1989})}\BibitemShut {NoStop}%
\bibitem [{\citenamefont {Redlich}(1984{\natexlab{a}})}]{PhysRevD.29.2366}%
  \BibitemOpen
  \bibfield  {author} {\bibinfo {author} {\bibfnamefont {A.~N.}\ \bibnamefont
  {Redlich}},\ }\href {\doibase 10.1103/PhysRevD.29.2366} {\bibfield  {journal}
  {\bibinfo  {journal} {Phys. Rev. D}\ }\textbf {\bibinfo {volume} {29}},\
  \bibinfo {pages} {2366} (\bibinfo {year} {1984}{\natexlab{a}})}\BibitemShut
  {NoStop}%
\bibitem [{\citenamefont {Redlich}(1984{\natexlab{b}})}]{Redlich:1983kn}%
  \BibitemOpen
  \bibfield  {author} {\bibinfo {author} {\bibfnamefont {A.~N.}\ \bibnamefont
  {Redlich}},\ }\href {\doibase 10.1103/PhysRevLett.52.18} {\bibfield
  {journal} {\bibinfo  {journal} {Phys. Rev. Lett.}\ }\textbf {\bibinfo
  {volume} {52}},\ \bibinfo {pages} {18} (\bibinfo {year}
  {1984}{\natexlab{b}})},\ \bibinfo {note} {[,364(1983)]}\BibitemShut {NoStop}%
%%CITATION = PRLTA,52,18;%%
\bibitem [{\citenamefont {Niemi}\ and\ \citenamefont
  {Semenoff}(1983)}]{PhysRevLett.51.2077}%
  \BibitemOpen
  \bibfield  {author} {\bibinfo {author} {\bibfnamefont {A.~J.}\ \bibnamefont
  {Niemi}}\ and\ \bibinfo {author} {\bibfnamefont {G.~W.}\ \bibnamefont
  {Semenoff}},\ }\href {\doibase 10.1103/PhysRevLett.51.2077} {\bibfield
  {journal} {\bibinfo  {journal} {Phys. Rev. Lett.}\ }\textbf {\bibinfo
  {volume} {51}},\ \bibinfo {pages} {2077} (\bibinfo {year}
  {1983})}\BibitemShut {NoStop}%
\bibitem [{\citenamefont {Karthik}\ and\ \citenamefont
  {Narayanan}(2018)}]{Karthik:2018tnh}%
  \BibitemOpen
  \bibfield  {author} {\bibinfo {author} {\bibfnamefont {N.}~\bibnamefont
  {Karthik}}\ and\ \bibinfo {author} {\bibfnamefont {R.}~\bibnamefont
  {Narayanan}},\ }\href {\doibase 10.1103/PhysRevLett.121.041602} {\bibfield
  {journal} {\bibinfo  {journal} {Phys. Rev. Lett.}\ }\textbf {\bibinfo
  {volume} {121}},\ \bibinfo {pages} {041602} (\bibinfo {year} {2018})},\
  \Eprint {http://arxiv.org/abs/1803.03596} {arXiv:1803.03596 [hep-lat]}
  \BibitemShut {NoStop}%
%%CITATION = ARXIV:1803.03596;%%
\bibitem [{\citenamefont {Karthik}\ and\ \citenamefont
  {Narayanan}(2016)}]{Karthik:2016ppr}%
  \BibitemOpen
  \bibfield  {author} {\bibinfo {author} {\bibfnamefont {N.}~\bibnamefont
  {Karthik}}\ and\ \bibinfo {author} {\bibfnamefont {R.}~\bibnamefont
  {Narayanan}},\ }\href {\doibase 10.1103/PhysRevD.94.065026} {\bibfield
  {journal} {\bibinfo  {journal} {Phys. Rev.}\ }\textbf {\bibinfo {volume}
  {D94}},\ \bibinfo {pages} {065026} (\bibinfo {year} {2016})},\ \Eprint
  {http://arxiv.org/abs/1606.04109} {arXiv:1606.04109 [hep-th]} \BibitemShut
  {NoStop}%
%%CITATION = ARXIV:1606.04109;%%
\bibitem [{\citenamefont {Lapa}(2019)}]{Lapa:2019fiv}%
  \BibitemOpen
  \bibfield  {author} {\bibinfo {author} {\bibfnamefont {M.~F.}\ \bibnamefont
  {Lapa}},\ }\href@noop {} {\  (\bibinfo {year} {2019})},\ \Eprint
  {http://arxiv.org/abs/1903.06719} {arXiv:1903.06719 [cond-mat.other]}
  \BibitemShut {NoStop}%
%%CITATION = ARXIV:1903.06719;%%
\bibitem [{\citenamefont {Kapustin}\ and\ \citenamefont
  {Thorngren}(2014)}]{Kapustin:2014lwa}%
  \BibitemOpen
  \bibfield  {author} {\bibinfo {author} {\bibfnamefont {A.}~\bibnamefont
  {Kapustin}}\ and\ \bibinfo {author} {\bibfnamefont {R.}~\bibnamefont
  {Thorngren}},\ }\href {\doibase 10.1103/PhysRevLett.112.231602} {\bibfield
  {journal} {\bibinfo  {journal} {Phys. Rev. Lett.}\ }\textbf {\bibinfo
  {volume} {112}},\ \bibinfo {pages} {231602} (\bibinfo {year} {2014})},\
  \Eprint {http://arxiv.org/abs/1403.0617} {arXiv:1403.0617 [hep-th]}
  \BibitemShut {NoStop}%
%%CITATION = ARXIV:1403.0617;%%
\bibitem [{\citenamefont {Deser}\ \emph {et~al.}(1997)\citenamefont {Deser},
  \citenamefont {Griguolo},\ and\ \citenamefont {Seminara}}]{Deser:1997nv}%
  \BibitemOpen
  \bibfield  {author} {\bibinfo {author} {\bibfnamefont {S.}~\bibnamefont
  {Deser}}, \bibinfo {author} {\bibfnamefont {L.}~\bibnamefont {Griguolo}}, \
  and\ \bibinfo {author} {\bibfnamefont {D.}~\bibnamefont {Seminara}},\ }\href
  {\doibase 10.1103/PhysRevLett.79.1976} {\bibfield  {journal} {\bibinfo
  {journal} {Phys. Rev. Lett.}\ }\textbf {\bibinfo {volume} {79}},\ \bibinfo
  {pages} {1976} (\bibinfo {year} {1997})},\ \Eprint
  {http://arxiv.org/abs/hep-th/9705052} {arXiv:hep-th/9705052 [hep-th]}
  \BibitemShut {NoStop}%
%%CITATION = HEP-TH/9705052;%%
\bibitem [{\citenamefont {Nogueira}\ and\ \citenamefont
  {Kleinert}(2005)}]{Nogueira:2005aj}%
  \BibitemOpen
  \bibfield  {author} {\bibinfo {author} {\bibfnamefont {F.~S.}\ \bibnamefont
  {Nogueira}}\ and\ \bibinfo {author} {\bibfnamefont {H.}~\bibnamefont
  {Kleinert}},\ }\href {\doibase 10.1103/PhysRevLett.95.176406} {\bibfield
  {journal} {\bibinfo  {journal} {Phys. Rev. Lett.}\ }\textbf {\bibinfo
  {volume} {95}},\ \bibinfo {pages} {176406} (\bibinfo {year} {2005})},\
  \Eprint {http://arxiv.org/abs/cond-mat/0501022} {arXiv:cond-mat/0501022
  [cond-mat]} \BibitemShut {NoStop}%
%%CITATION = COND-MAT/0501022;%%
\bibitem [{\citenamefont {Callan}\ and\ \citenamefont
  {Harvey}(1985)}]{Callan:1984sa}%
  \BibitemOpen
  \bibfield  {author} {\bibinfo {author} {\bibfnamefont {C.~G.}\ \bibnamefont
  {Callan}, \bibfnamefont {Jr.}}\ and\ \bibinfo {author} {\bibfnamefont
  {J.~A.}\ \bibnamefont {Harvey}},\ }\href {\doibase
  10.1016/0550-3213(85)90489-4} {\bibfield  {journal} {\bibinfo  {journal}
  {Nucl. Phys.}\ }\textbf {\bibinfo {volume} {B250}},\ \bibinfo {pages} {427}
  (\bibinfo {year} {1985})}\BibitemShut {NoStop}%
%%CITATION = NUPHA,B250,427;%%
\bibitem [{\citenamefont {Steinberg}\ and\ \citenamefont
  {Swingle}(2019)}]{Steinberg:2019uqb}%
  \BibitemOpen
  \bibfield  {author} {\bibinfo {author} {\bibfnamefont {J.}~\bibnamefont
  {Steinberg}}\ and\ \bibinfo {author} {\bibfnamefont {B.}~\bibnamefont
  {Swingle}},\ }\href@noop {} {\  (\bibinfo {year} {2019})},\ \Eprint
  {http://arxiv.org/abs/1901.04984} {arXiv:1901.04984 [cond-mat.str-el]}
  \BibitemShut {NoStop}%
\bibitem [{\citenamefont {Müller}\ \emph {et~al.}(2016)\citenamefont
  {Müller}, \citenamefont {Schlichting},\ and\ \citenamefont
  {Sharma}}]{Mueller:2016ven}%
  \BibitemOpen
  \bibfield  {author} {\bibinfo {author} {\bibfnamefont {N.}~\bibnamefont
  {Müller}}, \bibinfo {author} {\bibfnamefont {S.}~\bibnamefont
  {Schlichting}}, \ and\ \bibinfo {author} {\bibfnamefont {S.}~\bibnamefont
  {Sharma}},\ }\href {\doibase 10.1103/PhysRevLett.117.142301} {\bibfield
  {journal} {\bibinfo  {journal} {Phys. Rev. Lett.}\ }\textbf {\bibinfo
  {volume} {117}},\ \bibinfo {pages} {142301} (\bibinfo {year} {2016})},\
  \Eprint {http://arxiv.org/abs/1606.00342} {arXiv:1606.00342 [hep-ph]}
  \BibitemShut {NoStop}%
%%CITATION = ARXIV:1606.00342;%%
\bibitem [{\citenamefont {Spitz}\ and\ \citenamefont
  {Berges}(2019)}]{Spitz:2018eps}%
  \BibitemOpen
  \bibfield  {author} {\bibinfo {author} {\bibfnamefont {D.}~\bibnamefont
  {Spitz}}\ and\ \bibinfo {author} {\bibfnamefont {J.}~\bibnamefont {Berges}},\
  }\href {\doibase 10.1103/PhysRevD.99.036020} {\bibfield  {journal} {\bibinfo
  {journal} {Phys. Rev.}\ }\textbf {\bibinfo {volume} {D99}},\ \bibinfo {pages}
  {036020} (\bibinfo {year} {2019})},\ \Eprint
  {http://arxiv.org/abs/1812.05835} {arXiv:1812.05835 [hep-ph]} \BibitemShut
  {NoStop}%
%%CITATION = ARXIV:1812.05835;%%
\bibitem [{\citenamefont {Aarts}\ and\ \citenamefont
  {Smit}(1999)}]{Aarts:1998td}%
  \BibitemOpen
  \bibfield  {author} {\bibinfo {author} {\bibfnamefont {G.}~\bibnamefont
  {Aarts}}\ and\ \bibinfo {author} {\bibfnamefont {J.}~\bibnamefont {Smit}},\
  }\href {\doibase 10.1016/S0550-3213(99)00320-X} {\bibfield  {journal}
  {\bibinfo  {journal} {Nucl. Phys.}\ }\textbf {\bibinfo {volume} {B555}},\
  \bibinfo {pages} {355} (\bibinfo {year} {1999})},\ \Eprint
  {http://arxiv.org/abs/hep-ph/9812413} {arXiv:hep-ph/9812413 [hep-ph]}
  \BibitemShut {NoStop}%
%%CITATION = HEP-PH/9812413;%%
\bibitem [{\citenamefont {Berges}\ \emph {et~al.}(2011)\citenamefont {Berges},
  \citenamefont {Gelfand},\ and\ \citenamefont {Pruschke}}]{Berges:2010zv}%
  \BibitemOpen
  \bibfield  {author} {\bibinfo {author} {\bibfnamefont {J.}~\bibnamefont
  {Berges}}, \bibinfo {author} {\bibfnamefont {D.}~\bibnamefont {Gelfand}}, \
  and\ \bibinfo {author} {\bibfnamefont {J.}~\bibnamefont {Pruschke}},\ }\href
  {\doibase 10.1103/PhysRevLett.107.061301} {\bibfield  {journal} {\bibinfo
  {journal} {Phys. Rev. Lett.}\ }\textbf {\bibinfo {volume} {107}},\ \bibinfo
  {pages} {061301} (\bibinfo {year} {2011})},\ \Eprint
  {http://arxiv.org/abs/1012.4632} {arXiv:1012.4632 [hep-ph]} \BibitemShut
  {NoStop}%
%%CITATION = ARXIV:1012.4632;%%
\bibitem [{\citenamefont {Saffin}\ and\ \citenamefont
  {Tranberg}(2011)}]{Saffin:2011kc}%
  \BibitemOpen
  \bibfield  {author} {\bibinfo {author} {\bibfnamefont {P.~M.}\ \bibnamefont
  {Saffin}}\ and\ \bibinfo {author} {\bibfnamefont {A.}~\bibnamefont
  {Tranberg}},\ }\href {\doibase 10.1007/JHEP07(2011)066} {\bibfield  {journal}
  {\bibinfo  {journal} {JHEP}\ }\textbf {\bibinfo {volume} {07}},\ \bibinfo
  {pages} {066} (\bibinfo {year} {2011})},\ \Eprint
  {http://arxiv.org/abs/1105.5546} {arXiv:1105.5546 [hep-ph]} \BibitemShut
  {NoStop}%
%%CITATION = ARXIV:1105.5546;%%
\bibitem [{\citenamefont {Hebenstreit}\ \emph
  {et~al.}(2013{\natexlab{b}})\citenamefont {Hebenstreit}, \citenamefont
  {Berges},\ and\ \citenamefont {Gelfand}}]{Hebenstreit:2013qxa}%
  \BibitemOpen
  \bibfield  {author} {\bibinfo {author} {\bibfnamefont {F.}~\bibnamefont
  {Hebenstreit}}, \bibinfo {author} {\bibfnamefont {J.}~\bibnamefont {Berges}},
  \ and\ \bibinfo {author} {\bibfnamefont {D.}~\bibnamefont {Gelfand}},\ }\href
  {\doibase 10.1103/PhysRevD.87.105006} {\bibfield  {journal} {\bibinfo
  {journal} {Phys. Rev.}\ }\textbf {\bibinfo {volume} {D87}},\ \bibinfo {pages}
  {105006} (\bibinfo {year} {2013}{\natexlab{b}})},\ \Eprint
  {http://arxiv.org/abs/1302.5537} {arXiv:1302.5537 [hep-ph]} \BibitemShut
  {NoStop}%
%%CITATION = ARXIV:1302.5537;%%
\bibitem [{\citenamefont {Kasper}\ \emph {et~al.}(2014)\citenamefont {Kasper},
  \citenamefont {Hebenstreit},\ and\ \citenamefont {Berges}}]{Kasper:2014uaa}%
  \BibitemOpen
  \bibfield  {author} {\bibinfo {author} {\bibfnamefont {V.}~\bibnamefont
  {Kasper}}, \bibinfo {author} {\bibfnamefont {F.}~\bibnamefont {Hebenstreit}},
  \ and\ \bibinfo {author} {\bibfnamefont {J.}~\bibnamefont {Berges}},\ }\href
  {\doibase 10.1103/PhysRevD.90.025016} {\bibfield  {journal} {\bibinfo
  {journal} {Phys. Rev.}\ }\textbf {\bibinfo {volume} {D90}},\ \bibinfo {pages}
  {025016} (\bibinfo {year} {2014})},\ \Eprint {http://arxiv.org/abs/1403.4849}
  {arXiv:1403.4849 [hep-ph]} \BibitemShut {NoStop}%
%%CITATION = ARXIV:1403.4849;%%
\bibitem [{\citenamefont {Buividovich}\ and\ \citenamefont
  {Ulybyshev}(2016)}]{Buividovich:2015jfa}%
  \BibitemOpen
  \bibfield  {author} {\bibinfo {author} {\bibfnamefont {P.~V.}\ \bibnamefont
  {Buividovich}}\ and\ \bibinfo {author} {\bibfnamefont {M.~V.}\ \bibnamefont
  {Ulybyshev}},\ }\href {\doibase 10.1103/PhysRevD.94.025009} {\bibfield
  {journal} {\bibinfo  {journal} {Phys. Rev.}\ }\textbf {\bibinfo {volume}
  {D94}},\ \bibinfo {pages} {025009} (\bibinfo {year} {2016})},\ \Eprint
  {http://arxiv.org/abs/1509.02076} {arXiv:1509.02076 [hep-th]} \BibitemShut
  {NoStop}%
%%CITATION = ARXIV:1509.02076;%%
\bibitem [{\citenamefont {Gelfand}\ \emph {et~al.}(2016)\citenamefont
  {Gelfand}, \citenamefont {Hebenstreit},\ and\ \citenamefont
  {Berges}}]{Gelfand:2016prm}%
  \BibitemOpen
  \bibfield  {author} {\bibinfo {author} {\bibfnamefont {D.}~\bibnamefont
  {Gelfand}}, \bibinfo {author} {\bibfnamefont {F.}~\bibnamefont
  {Hebenstreit}}, \ and\ \bibinfo {author} {\bibfnamefont {J.}~\bibnamefont
  {Berges}},\ }\href {\doibase 10.1103/PhysRevD.93.085001} {\bibfield
  {journal} {\bibinfo  {journal} {Phys. Rev.}\ }\textbf {\bibinfo {volume}
  {D93}},\ \bibinfo {pages} {085001} (\bibinfo {year} {2016})},\ \Eprint
  {http://arxiv.org/abs/1601.03576} {arXiv:1601.03576 [hep-ph]} \BibitemShut
  {NoStop}%
%%CITATION = ARXIV:1601.03576;%%
\bibitem [{\citenamefont {Tanji}\ \emph {et~al.}(2016)\citenamefont {Tanji},
  \citenamefont {Mueller},\ and\ \citenamefont {Berges}}]{Tanji:2016dka}%
  \BibitemOpen
  \bibfield  {author} {\bibinfo {author} {\bibfnamefont {N.}~\bibnamefont
  {Tanji}}, \bibinfo {author} {\bibfnamefont {N.}~\bibnamefont {Mueller}}, \
  and\ \bibinfo {author} {\bibfnamefont {J.}~\bibnamefont {Berges}},\ }\href
  {\doibase 10.1103/PhysRevD.93.074507} {\bibfield  {journal} {\bibinfo
  {journal} {Phys. Rev.}\ }\textbf {\bibinfo {volume} {D93}},\ \bibinfo {pages}
  {074507} (\bibinfo {year} {2016})},\ \Eprint
  {http://arxiv.org/abs/1603.03331} {arXiv:1603.03331 [hep-ph]} \BibitemShut
  {NoStop}%
%%CITATION = ARXIV:1603.03331;%%
\bibitem [{\citenamefont {Mueller}\ \emph {et~al.}(2016)\citenamefont
  {Mueller}, \citenamefont {Hebenstreit},\ and\ \citenamefont
  {Berges}}]{Mueller:2016aao}%
  \BibitemOpen
  \bibfield  {author} {\bibinfo {author} {\bibfnamefont {N.}~\bibnamefont
  {Mueller}}, \bibinfo {author} {\bibfnamefont {F.}~\bibnamefont
  {Hebenstreit}}, \ and\ \bibinfo {author} {\bibfnamefont {J.}~\bibnamefont
  {Berges}},\ }\href {\doibase 10.1103/PhysRevLett.117.061601} {\bibfield
  {journal} {\bibinfo  {journal} {Phys. Rev. Lett.}\ }\textbf {\bibinfo
  {volume} {117}},\ \bibinfo {pages} {061601} (\bibinfo {year} {2016})},\
  \Eprint {http://arxiv.org/abs/1605.01413} {arXiv:1605.01413 [hep-ph]}
  \BibitemShut {NoStop}%
%%CITATION = ARXIV:1605.01413;%%
\bibitem [{\citenamefont {Hebenstreit}(2011)}]{Hebenstreit:2011pm}%
  \BibitemOpen
  \bibfield  {author} {\bibinfo {author} {\bibfnamefont {F.}~\bibnamefont
  {Hebenstreit}},\ }\emph {\bibinfo {title} {{Schwinger effect in inhomogeneous
  electric fields}}},\ \href@noop {} {Ph.D. thesis},\ \bibinfo  {school} {Graz
  U.} (\bibinfo {year} {2011}),\ \Eprint {http://arxiv.org/abs/1106.5965}
  {arXiv:1106.5965 [hep-ph]} \BibitemShut {NoStop}%
%%CITATION = ARXIV:1106.5965;%%
\bibitem [{\citenamefont {Pauli}\ and\ \citenamefont
  {Villars}(1949)}]{RevModPhys.21.434}%
  \BibitemOpen
  \bibfield  {author} {\bibinfo {author} {\bibfnamefont {W.}~\bibnamefont
  {Pauli}}\ and\ \bibinfo {author} {\bibfnamefont {F.}~\bibnamefont
  {Villars}},\ }\href {\doibase 10.1103/RevModPhys.21.434} {\bibfield
  {journal} {\bibinfo  {journal} {Rev. Mod. Phys.}\ }\textbf {\bibinfo {volume}
  {21}},\ \bibinfo {pages} {434} (\bibinfo {year} {1949})}\BibitemShut
  {NoStop}%
\bibitem [{\citenamefont {Kogut}\ and\ \citenamefont
  {Susskind}(1975)}]{Kogut:1974ag}%
  \BibitemOpen
  \bibfield  {author} {\bibinfo {author} {\bibfnamefont {J.~B.}\ \bibnamefont
  {Kogut}}\ and\ \bibinfo {author} {\bibfnamefont {L.}~\bibnamefont
  {Susskind}},\ }\href {\doibase 10.1103/PhysRevD.11.395} {\bibfield  {journal}
  {\bibinfo  {journal} {Phys. Rev.}\ }\textbf {\bibinfo {volume} {D11}},\
  \bibinfo {pages} {395} (\bibinfo {year} {1975})}\BibitemShut {NoStop}%
%%CITATION = PHRVA,D11,395;%%
\bibitem [{\citenamefont {Mace}\ \emph {et~al.}(2017)\citenamefont {Mace},
  \citenamefont {Mueller}, \citenamefont {Schlichting},\ and\ \citenamefont
  {Sharma}}]{Mace:2016shq}%
  \BibitemOpen
  \bibfield  {author} {\bibinfo {author} {\bibfnamefont {M.}~\bibnamefont
  {Mace}}, \bibinfo {author} {\bibfnamefont {N.}~\bibnamefont {Mueller}},
  \bibinfo {author} {\bibfnamefont {S.}~\bibnamefont {Schlichting}}, \ and\
  \bibinfo {author} {\bibfnamefont {S.}~\bibnamefont {Sharma}},\ }\href
  {\doibase 10.1103/PhysRevD.95.036023} {\bibfield  {journal} {\bibinfo
  {journal} {Phys. Rev.}\ }\textbf {\bibinfo {volume} {D95}},\ \bibinfo {pages}
  {036023} (\bibinfo {year} {2017})},\ \Eprint
  {http://arxiv.org/abs/1612.02477} {arXiv:1612.02477 [hep-lat]} \BibitemShut
  {NoStop}%
%%CITATION = ARXIV:1612.02477;%%
\bibitem [{\citenamefont {Zache}\ \emph {et~al.}(2019)\citenamefont {Zache},
  \citenamefont {Mueller}, \citenamefont {Schneider}, \citenamefont
  {Jendrzejewski}, \citenamefont {Berges},\ and\ \citenamefont
  {Hauke}}]{Zache:2018cqq}%
  \BibitemOpen
  \bibfield  {author} {\bibinfo {author} {\bibfnamefont {T.~V.}\ \bibnamefont
  {Zache}}, \bibinfo {author} {\bibfnamefont {N.}~\bibnamefont {Mueller}},
  \bibinfo {author} {\bibfnamefont {J.~T.}\ \bibnamefont {Schneider}}, \bibinfo
  {author} {\bibfnamefont {F.}~\bibnamefont {Jendrzejewski}}, \bibinfo {author}
  {\bibfnamefont {J.}~\bibnamefont {Berges}}, \ and\ \bibinfo {author}
  {\bibfnamefont {P.}~\bibnamefont {Hauke}},\ }\href {\doibase
  10.1103/PhysRevLett.122.050403} {\bibfield  {journal} {\bibinfo  {journal}
  {Phys. Rev. Lett.}\ }\textbf {\bibinfo {volume} {122}},\ \bibinfo {pages}
  {050403} (\bibinfo {year} {2019})},\ \Eprint
  {http://arxiv.org/abs/1808.07885} {arXiv:1808.07885 [quant-ph]} \BibitemShut
  {NoStop}%
%%CITATION = ARXIV:1808.07885;%%
\bibitem [{\citenamefont {Klco}\ \emph {et~al.}(2018)\citenamefont {Klco},
  \citenamefont {Dumitrescu}, \citenamefont {McCaskey}, \citenamefont {Morris},
  \citenamefont {Pooser}, \citenamefont {Sanz}, \citenamefont {Solano},
  \citenamefont {Lougovski},\ and\ \citenamefont {Savage}}]{Klco:2018kyo}%
  \BibitemOpen
  \bibfield  {author} {\bibinfo {author} {\bibfnamefont {N.}~\bibnamefont
  {Klco}}, \bibinfo {author} {\bibfnamefont {E.~F.}\ \bibnamefont
  {Dumitrescu}}, \bibinfo {author} {\bibfnamefont {A.~J.}\ \bibnamefont
  {McCaskey}}, \bibinfo {author} {\bibfnamefont {T.~D.}\ \bibnamefont
  {Morris}}, \bibinfo {author} {\bibfnamefont {R.~C.}\ \bibnamefont {Pooser}},
  \bibinfo {author} {\bibfnamefont {M.}~\bibnamefont {Sanz}}, \bibinfo {author}
  {\bibfnamefont {E.}~\bibnamefont {Solano}}, \bibinfo {author} {\bibfnamefont
  {P.}~\bibnamefont {Lougovski}}, \ and\ \bibinfo {author} {\bibfnamefont
  {M.~J.}\ \bibnamefont {Savage}},\ }\href {\doibase
  10.1103/PhysRevA.98.032331} {\bibfield  {journal} {\bibinfo  {journal} {Phys.
  Rev.}\ }\textbf {\bibinfo {volume} {A98}},\ \bibinfo {pages} {032331}
  (\bibinfo {year} {2018})},\ \Eprint {http://arxiv.org/abs/1803.03326}
  {arXiv:1803.03326 [quant-ph]} \BibitemShut {NoStop}%
%%CITATION = ARXIV:1803.03326;%%
\bibitem [{\citenamefont {Schweizer}\ \emph {et~al.}(2019)\citenamefont
  {Schweizer}, \citenamefont {Grusdt}, \citenamefont {Berngruber},
  \citenamefont {Barbiero}, \citenamefont {Demler}, \citenamefont {Goldman},
  \citenamefont {Bloch},\ and\ \citenamefont
  {Aidelsburger}}]{schweizer2019floquet}%
  \BibitemOpen
  \bibfield  {author} {\bibinfo {author} {\bibfnamefont {C.}~\bibnamefont
  {Schweizer}}, \bibinfo {author} {\bibfnamefont {F.}~\bibnamefont {Grusdt}},
  \bibinfo {author} {\bibfnamefont {M.}~\bibnamefont {Berngruber}}, \bibinfo
  {author} {\bibfnamefont {L.}~\bibnamefont {Barbiero}}, \bibinfo {author}
  {\bibfnamefont {E.}~\bibnamefont {Demler}}, \bibinfo {author} {\bibfnamefont
  {N.}~\bibnamefont {Goldman}}, \bibinfo {author} {\bibfnamefont
  {I.}~\bibnamefont {Bloch}}, \ and\ \bibinfo {author} {\bibfnamefont
  {M.}~\bibnamefont {Aidelsburger}},\ }\href@noop {} {\bibfield  {journal}
  {\bibinfo  {journal} {arXiv preprint arXiv:1901.07103}\ } (\bibinfo {year}
  {2019})}\BibitemShut {NoStop}%
\bibitem [{\citenamefont {Zohar}\ \emph {et~al.}(2013)\citenamefont {Zohar},
  \citenamefont {Cirac},\ and\ \citenamefont {Reznik}}]{Zohar:2013zla}%
  \BibitemOpen
  \bibfield  {author} {\bibinfo {author} {\bibfnamefont {E.}~\bibnamefont
  {Zohar}}, \bibinfo {author} {\bibfnamefont {J.~I.}\ \bibnamefont {Cirac}}, \
  and\ \bibinfo {author} {\bibfnamefont {B.}~\bibnamefont {Reznik}},\ }\href
  {\doibase 10.1103/PhysRevA.88.023617} {\bibfield  {journal} {\bibinfo
  {journal} {Phys. Rev.}\ }\textbf {\bibinfo {volume} {A88}},\ \bibinfo {pages}
  {023617} (\bibinfo {year} {2013})},\ \Eprint {http://arxiv.org/abs/1303.5040}
  {arXiv:1303.5040 [quant-ph]} \BibitemShut {NoStop}%
%%CITATION = ARXIV:1303.5040;%%
\bibitem [{\citenamefont {Bender}\ \emph {et~al.}(2018)\citenamefont {Bender},
  \citenamefont {Zohar}, \citenamefont {Farace},\ and\ \citenamefont
  {Cirac}}]{Bender:2018rdp}%
  \BibitemOpen
  \bibfield  {author} {\bibinfo {author} {\bibfnamefont {J.}~\bibnamefont
  {Bender}}, \bibinfo {author} {\bibfnamefont {E.}~\bibnamefont {Zohar}},
  \bibinfo {author} {\bibfnamefont {A.}~\bibnamefont {Farace}}, \ and\ \bibinfo
  {author} {\bibfnamefont {J.~I.}\ \bibnamefont {Cirac}},\ }\href {\doibase
  10.1088/1367-2630/aadb71} {\bibfield  {journal} {\bibinfo  {journal} {New J.
  Phys.}\ }\textbf {\bibinfo {volume} {20}},\ \bibinfo {pages} {093001}
  (\bibinfo {year} {2018})},\ \Eprint {http://arxiv.org/abs/1804.02082}
  {arXiv:1804.02082 [quant-ph]} \BibitemShut {NoStop}%
%%CITATION = ARXIV:1804.02082;%%
\bibitem [{\citenamefont {Ambjorn}\ \emph {et~al.}(1983)\citenamefont
  {Ambjorn}, \citenamefont {Greensite},\ and\ \citenamefont
  {Peterson}}]{Ambjorn:1983hp}%
  \BibitemOpen
  \bibfield  {author} {\bibinfo {author} {\bibfnamefont {J.}~\bibnamefont
  {Ambjorn}}, \bibinfo {author} {\bibfnamefont {J.}~\bibnamefont {Greensite}},
  \ and\ \bibinfo {author} {\bibfnamefont {C.}~\bibnamefont {Peterson}},\
  }\href {\doibase 10.1016/0550-3213(83)90585-0} {\bibfield  {journal}
  {\bibinfo  {journal} {Nucl. Phys.}\ }\textbf {\bibinfo {volume} {B221}},\
  \bibinfo {pages} {381} (\bibinfo {year} {1983})}\BibitemShut {NoStop}%
%%CITATION = NUPHA,B221,381;%%
\bibitem [{\citenamefont {Hebenstreit}\ \emph {et~al.}(2009)\citenamefont
  {Hebenstreit}, \citenamefont {Alkofer}, \citenamefont {Dunne},\ and\
  \citenamefont {Gies}}]{Hebenstreit:2009km}%
  \BibitemOpen
  \bibfield  {author} {\bibinfo {author} {\bibfnamefont {F.}~\bibnamefont
  {Hebenstreit}}, \bibinfo {author} {\bibfnamefont {R.}~\bibnamefont
  {Alkofer}}, \bibinfo {author} {\bibfnamefont {G.~V.}\ \bibnamefont {Dunne}},
  \ and\ \bibinfo {author} {\bibfnamefont {H.}~\bibnamefont {Gies}},\ }\href
  {\doibase 10.1103/PhysRevLett.102.150404} {\bibfield  {journal} {\bibinfo
  {journal} {Phys. Rev. Lett.}\ }\textbf {\bibinfo {volume} {102}},\ \bibinfo
  {pages} {150404} (\bibinfo {year} {2009})},\ \Eprint
  {http://arxiv.org/abs/0901.2631} {arXiv:0901.2631 [hep-ph]} \BibitemShut
  {NoStop}%
%%CITATION = ARXIV:0901.2631;%%
\bibitem [{\citenamefont {Dumlu}\ and\ \citenamefont
  {Dunne}(2010)}]{Dumlu:2010ua}%
  \BibitemOpen
  \bibfield  {author} {\bibinfo {author} {\bibfnamefont {C.~K.}\ \bibnamefont
  {Dumlu}}\ and\ \bibinfo {author} {\bibfnamefont {G.~V.}\ \bibnamefont
  {Dunne}},\ }\href {\doibase 10.1103/PhysRevLett.104.250402} {\bibfield
  {journal} {\bibinfo  {journal} {Phys. Rev. Lett.}\ }\textbf {\bibinfo
  {volume} {104}},\ \bibinfo {pages} {250402} (\bibinfo {year} {2010})},\
  \Eprint {http://arxiv.org/abs/1004.2509} {arXiv:1004.2509 [hep-th]}
  \BibitemShut {NoStop}%
%%CITATION = ARXIV:1004.2509;%%
\end{thebibliography}%

\end{document}